\documentclass[twocolumn,showpacs,showkeys,amssymb,nobibnotes,superscriptaddress,aps,prx,bibnotes]{revtex4-2}

\usepackage{graphicx}
\usepackage{dcolumn}
\usepackage{bm}
\usepackage[T1]{fontenc}
\usepackage{mathptmx}
\usepackage{mathrsfs}
\usepackage{physics}
\usepackage{simplewick}
\usepackage{amsthm,amsmath,amssymb}

\usepackage[bottom]{footmisc}

\usepackage{amssymb}
\usepackage{amsmath}
\usepackage{amsfonts}
\usepackage{bm}
\usepackage{graphicx}
\usepackage{subfigure}
\usepackage[dvipsnames]{xcolor}
\usepackage{calc}
\usepackage{epsfig}
\usepackage{color}
\usepackage[colorlinks,citecolor=blue]{hyperref}
\begin{document}
\bibliographystyle{apsrev4-1}
\newcommand{\be}{\begin{equation}}
\newcommand{\ee}{\end{equation}}
\newcommand{\bs}{\begin{split}}
\newcommand{\es}{\end{split}}
\newcommand{\R}[1]{\textcolor{red}{#1}}
\newcommand{\B}[1]{\textcolor{blue}{#1}}

\title{Semiclassical gravity phenomenology under the causal-conditional quantum measurement prescription II: Heisenberg picture and apparent optical entanglement}
\author{Yubao Liu}
\affiliation{Center for Gravitational Experiment, Hubei Key Laboratory of Gravitation and Quantum Physics, School of Physics, Huazhong University of Science and Technology, Wuhan, 430074, China}
\author{Wenjie Zhong}
\affiliation{Center for Gravitational Experiment, Hubei Key Laboratory of Gravitation and Quantum Physics, School of Physics, Huazhong University of Science and Technology, Wuhan, 430074, China}
\author{Yanbei Chen}
\email{yanbei@caltech.edu}
\affiliation{Burke Institute of Theoretical Physics, California Institute of Technology, Pasadena, CA, 91125, USA}
\author{Yiqiu Ma}
\email{myqphy@hust.edu.cn}
\affiliation{Center for Gravitational Experiment, Hubei Key Laboratory of Gravitation and Quantum Physics, School of Physics, Huazhong University of Science and Technology, Wuhan, 430074, China}
\affiliation{Department of Astronomy, School of Physics, Huazhong University of Science and Technology, Wuhan, 430074, China}

\begin{abstract}
The evolution of quantum states influenced by semiclassical gravity is distinct from that in quantum gravity theory due to the presence of a state-dependent gravitational potential. This state-dependent potential introduces nonlinearity into the state evolution, of which the theory is named Schr{\"o}dinger-Newton\,(SN) theory. The formalism for understanding the continuous quantum measurement process on the quantum state in the context of semiclassical gravity theory has been previously discussed using the Schr{\"o}dinger picture in Paper I\,\cite{Liu2023}.  In this work, an equivalent formalism using the Heisenberg picture is developed and applied to the analysis of two optomechanical experiment protocols that targeted testing the quantum nature of gravity. This Heisenberg picture formalism of the SN theory has the advantage of helping the investigation of the covariance matrices of the outgoing light fields in these protocols and further the entanglement features. We found that the classical gravity between the quantum trajectories of two mirrors under continuous quantum measurement in the SN theory can induce an apparent entanglement of the outgoing light field\,(though there is no quantum entanglement of the mirrors), which could serve as a false alarm for those experiments designed for probing the quantum gravity induced entanglement.
\end{abstract}

\maketitle
\section{Introduction}
In the quantum era, advancements in modern quantum technology have paved the way for experimental investigations into the characteristics of gravitational fields generated by quantum objects\,\cite{Aspelmyer2014,Chen_2013,Aspelmyer2012,Hoang2016,Mason2019,Rossi2018,Jayich2008,Yu2020,Aggarwal2020,Cata2020,Ando2010,ubhi2022active}. The initial step in these explorations involves conducting experiments that can assess the quantum aspects of gravitational fields\,\cite{Howl2018,Carney_2019,Lami2024,Belenchia2018}. Generally, a gravitational field comprises a near-zone Newtonian component (Newtonian gravity) and transverse wave components (gravitational waves). Several studies present compelling arguments suggesting that demonstrating the quantum nature of Newtonian gravity strongly indicates the quantum nature of gravitational waves\,\cite{Bose2017,Miao2020,krisnanda2020}. Consequently, these arguments encourage us to concentrate on testing the quantum characteristics of Newtonian gravity, which is substantially more feasible than directly probing the quantum nature of gravitational waves.

The fundamental design principle of these experiments is to identify observational signatures that can differentiate between quantum gravity and semi-classical gravity theories\,\cite{Yang2013,Helou2017,Grossardt2016,Gan2016,Datta_2021,Miao2020}. These two theoretical frameworks are distinguished by their treatment of the gravitational field: quantum gravity posits that the gravitational field represents a quantum degree of freedom (d.o.f.) capable of existing in superposition states and transmitting quantum information\,\cite{Bose2017,krisnanda2020,Miao2020,Marshman2020,Danielson2022,Christodoulou2023,Nimmrichter2015}. Accordingly, the joint quantum state of gravitational and matter degrees of freedom evolves \emph{linearly} within their \emph{joint Hilbert space}. In contrast, the semi-classical gravity theory maintains the classical spacetime curvature $G_{\mu\nu}$, which is sourced by the quantum expectation value of the matter's energy-momentum tensor, $\langle \psi|\hat{T}_{\mu\nu}|\psi\rangle$, where $|\psi\rangle$ represents the quantum state of the matter\,\cite{Mueller1962,Rosenfeld1963}. In other words, in the context of semi-classical gravity, the gravitational potential is dependent on the state of the system. Notably, in the non-relativistic limit, such a \emph{state-dependent gravitational potential} influences the evolution of $|\psi\rangle$, resulting in a \emph{nonlinear} evolution as described by the \emph{Schr{\"o}dinger-Newton (SN) equation}\,\cite{Diosi1989,Penrose1996,Diosi1998,Carlip_2008,Bahrami_2014,Bassi_2017}. Consequently, by examining the gravity-induced quantum state evolution, one can extract observational signatures that distinguish between quantum and classical gravity interactions\,\cite{oppenheim2023,Oppenheim2023A,Page1981,bose2023M,Kafri_2015,Carney2021,Kafri_2014,Tilloy2016}\footnote{It is essential to remember that these gravitational interactions must be sourced by a quantum state.}.

An intriguing and promising type of experiment involves exploring the quantum nature of gravity by examining quantum correlations and entanglements generated by \emph{quantum Newtonian gravity}, as proposed in \,\cite{Bose2017,Miao2020,krisnanda2020,Matsumura2022,Miki2024}. Experimental protocols for testing these gravity-induced entanglements (GIE) include gravitationally coupled spin systems\,\cite{Bose2017} and mechanical oscillator systems\,\cite{Miao2020}. In the latter case, the GIE is probed via an optical field that couples with the mechanical oscillators. Essentially, the GIE of the two mechanical oscillators is mapped onto two optical fields, creating correlations and entanglement between the optical fields. On the contrary, in the Schr{\"o}dinger-Newton (SN) scenario where mutual gravity is classical, the two mechanical oscillators cannot establish entanglement through a classical channel. Thus, previous work \,\cite{Miao2020} concludes that there will be no correlations or entanglement in the optical fields in the SN theory. Therefore, the presence of correlation or entanglement in the outgoing light fields serves as a definitive indicator of quantum gravity.

However, this result, while insightful, requires careful examination, as we demonstrated in our previous work\,\cite{Liu2023}, referred to as Paper I herein. Paper I introduced a "causal-conditional formalism" to address the process of continuous quantum measurement of a mechanical quantum state in the SN theory. The findings reveal that classical gravity generated by the quantum trajectories\,\cite{Chen_2013,Rossi2019} of two oscillator mirrors can also induce a \emph{correlation} in the outgoing light field, which is nearly indistinguishable from that induced by quantum gravity. These quantum trajectories result from the continuous quantum measurement of both mirrors.  

The correlation of light identified in Paper I does not imply a quantum correlation between the mirrors, as classical gravity cannot transmit quantum information\,\cite{Bennett1996,Griffiths2007,Galindo2002}. Consequently, this work demonstrates that examining the statistical properties of light might mistakenly suggest the presence of a quantum correlation between mirrors, potentially resulting in a false indication of quantum gravity. However, in the context of GIE-experiments, our primary interest lies in determining whether the SN theory could similarly create a misleading signal when the light field's entanglement, which represents a specific form of correlation, is assessed. Within the SN framework, analyzing the entanglement structure of outgoing optical fields necessitates new theoretical approaches.

As discussed in Section\,\ref{sec:master_equation_method}, the conditional density matrix approach we developed in Paper I, based on the Schrödinger picture, has limitations. It cannot compute the covariance matrix of the outgoing light and is thus unsuitable for investigating the entanglement structure. Consequently, Paper I does not determine whether this light correlation, induced by the classical-gravity-coupled quantum trajectories, implies a false alarm of GIE. In standard quantum mechanics, the logarithmic negativity\,\cite{Simon2000,Duan2000,Horodecki2009} quantitatively measures the entanglement of a joint quantum Gaussian state, determined by the covariance matrix of the outgoing light. To address the question of correlation nature, we require an alternative approach in the \emph{Heisenberg picture} to handle the continuous quantum measurement of oscillator mirrors in the SN theory and calculate this covariance matrix. Methodologically, this work will develop such an approach, demonstrating its equivalency to the conditional density matrix method.

Addressing this question is crucial for testing the quantum nature of gravity. Calculations in this paper will show that the quantum correlation of outgoing lights induced by classical gravity indeed constitutes an entanglement-like structure\,(named apparent entanglement), closely resembling the optical entanglement induced by quantum gravity. (false alarm, mirror no entanglement). Careful interpretation of GIE experimental results involving quantum measurement is essential.

The structure of this paper is organized as follows.  In Section\,\ref{sec2}, a review of the main results obtained in Paper I is presented, including an introduction to the setup of the experimental protocol. Section\,\ref{sec3} is devoted to the development of the method based on the Heisenberg picture, which allows us to compute the covariance matrix of the outgoing light. The equivalency of this Heisenberg-picture method to the conditional master equation method will be explicitly presented. Section\,\ref{sec5} will apply the method developed in Section\,\ref{sec3} to the optomechanical protocol for detecting gravity-induced entanglement. The physical interpretation of the result will be thoroughly discussed in Section\,\ref{sec5} and finally we will conclude the paper.

\section{Optomechanical system affected by semi-classical gravity}\label{sec2}
Before introducing the Heisenberg-picture method in SN theory, we will outline the optomechanical protocols designed to test the quantum nature of gravity .

In Paper I, two optomechanical protocols are discussed, termed the "self-gravity protocol" and "mutual gravity protocol" (see Fig. \ref{fig:schemes}). Although the optical entanglement issue in this paper arises from the mutual gravity protocol, the first protocol is essential for developing and illustrating the Heisenberg-picture method.

\begin{figure}[h]
\centering
\includegraphics[width=0.45\textwidth]{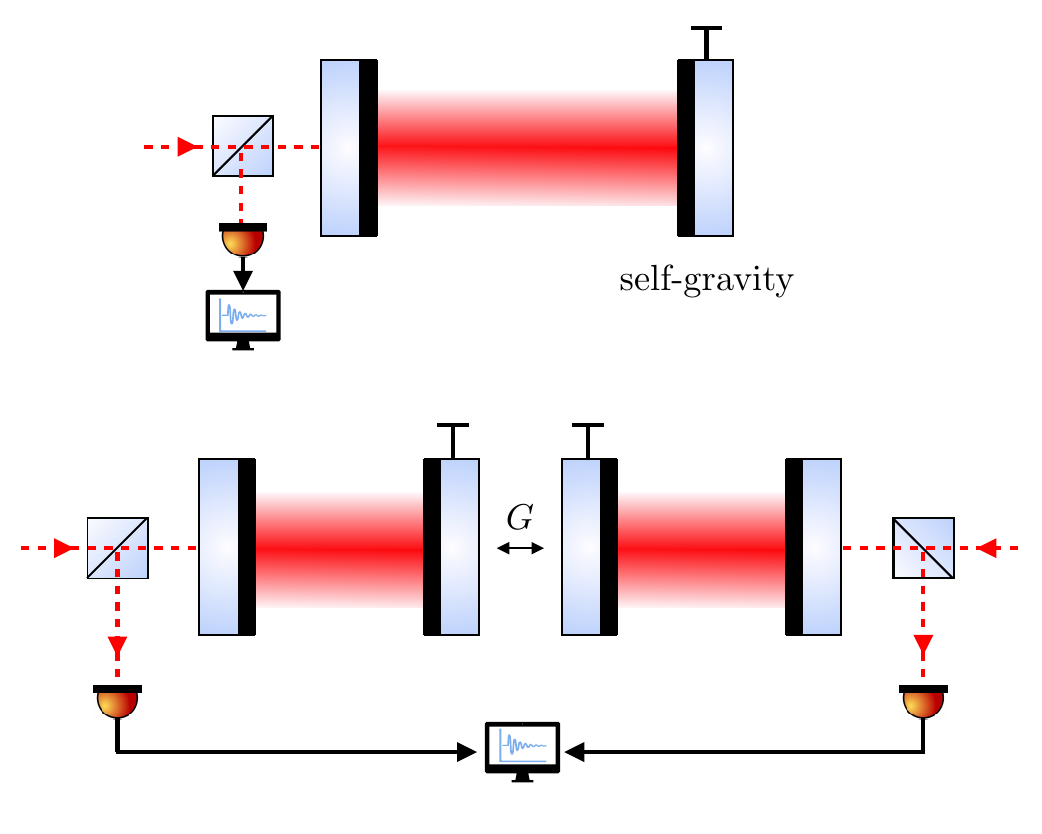}
\caption{Optomechanical protocol for testing SN theory. Upper panel: self gravity protocol, lower panel: mutual gravity protocol.}\label{fig:schemes}
\end{figure}

The self-gravity protocol is an optomechanical system affected by the mirror's self-gravity field. When gravity is classical, this self-gravity potential depends on the quantum state of the mirror's center-of-mass \,(CoM) motion\,\cite{Yang2013}. Information about the mirror's CoM motion is transferred into the light field through optomechanical coupling. The SN Hamiltonian is given by: 
\begin{equation}\label{eq:Hamiltonian_self_sn}
\hat{H}_{\rm self,\rm SN}=\frac{\hat{p}^2}{2M}+\frac{1}{2}M\omega_m^2\hat{x}^2+\frac{1}{2}M\omega^2_{\rm SN}(\hat{x}-\langle\hat{x}\rangle)^2+\hbar\alpha\hat{a}_1\hat{x},
\end{equation}

where $\alpha$ represents the optomechanical measurement strength and $\hat a_1$ is the amplitude quadrature operator of the light field. The SN term describes the state-dependent self-gravity effect, with the SN frequency\,\cite{Yang2013} $\omega_{\rm SN}=\sqrt{Gm/6\sqrt{\pi}x_{\rm zp}^3}$, with the $m,x_{\rm zp}$ are the ion mass and zero-point displacement of the ion oscillates in the crystal field, respectively. The first two terms describe the Hamiltonian of a free mechanical oscillator without gravity effects. When gravity is quantum, the self-gravity effect is negligible, rendering the SN term absent\,\cite{Yang2013}:
\begin{equation}\label{eq:Hamiltonian_self_qg}
\hat{H}_{\rm self,\rm QG}=\frac{\hat{p}^2}{2M}+\frac{1}{2}M\omega_m^2\hat{x}^2+\hbar\alpha\hat{a}_1\hat{x}+o(G^2),
\end{equation}

The mutual gravity protocol was introduced in \cite{Miao2020}, where two mechanical oscillators\,(A,B) are coupled via their mutual gravity. In the classical gravity scenario, the corresponding Hamiltonian\,(mechanical part)is:
\be\label{eq:Hamiltonian_mutual_sn}
\begin{split}
\hat H_{m,\rm SN}=\sum_{A/B}\left[\frac{\hat p^2_{A/B}}{2M}+\frac{M\omega_m^2}{2}\hat x^2_{A/B}-\mathcal{C}(\hat x_{A/B}-\langle\hat x_{B/A}\rangle)^2\right],
\end{split}
\ee
where $\mathcal{C}=M\Lambda G\rho$, with $\Lambda,\rho$ being the geometric factor and mass density of the oscillator mirrors, respectively. Under quantum gravity, the Hamiltonian is modified to:

\be\label{eq:Hamiltonian_mutual_qg}
\begin{split}
\hat H_{m,\rm QG}=\sum_{A/B}\left[\frac{\hat p^2_{A/B}}{2M}+\frac{M\omega_m^2}{2}\hat x^2_{A/B}-\mathcal{C}(\hat x_{A/B}-\hat x_{B/A})^2\right].
\end{split}
\ee
Two light fields independently couple to the mechanical oscillators as follows:
\be
\hat H_{\rm int}=-\hbar\alpha(\hat a_1\hat x_A+\hat b_1\hat x_B),
\ee
assuming both light fields have the same coherent amplitude $\alpha$. 

In these two protocols, the motion of the test mass mirrors is probed through their continuous coupling with the optical field. When the phase quadrature of the outgoing field is measured, it collapses the opto-mechanical joint quantum state. This collapse leaves the mirror's quantum trajectory, which can have a significant impact on the SN dynamics.

\section{Causal Conditional dynamics in the Schr{\"o}dinger picture}\label{sec:master_equation_method}
This section provides an overview of the conditional master equation method as discussed in Paper I, elucidating why this approach is inadequate for analyzing gravity-induced optical entanglement. Additionally, the result in Paper I is re-derived from the filtering problem perspective. This re-derivation will be instrumental in demonstrating the equivalence between the Heisenberg-picture method and the conditional master equation method.
\subsection{Causal conditional approach to continuous quantum measurement}
Consider the optomechanical system as an example of a continuous quantum measurement process\,\cite{Doherty1999,Hopkins2003,Ebhardt2009,Chen_2013}. During each infinitesimal temporal interval, entanglement is established between the incoming light fields and the mechanical oscillator, followed by a projective measurement of the outgoing light fields. This projective optical measurement collapses the joint quantum state into the Hilbert space of the mechanical oscillator, thereby generating a conditional mechanical state. This process occurs continuously, with the mean position and momentum of the conditional mechanical state tracing a \emph{quantum trajectory} in phase space\,\cite{Chen_2013,Rossi2019}. This is referred to as the \emph{causal conditional formalism}, which characterizes continuous quantum measurement processes. It is important to note that this causal conditional formalism is entirely equivalent to the pre-selection and post-selection formalism in standard quantum mechanics \,\cite{Aharonov1964}.

\begin{figure}[h]
\centering
\includegraphics[width=0.5\textwidth]{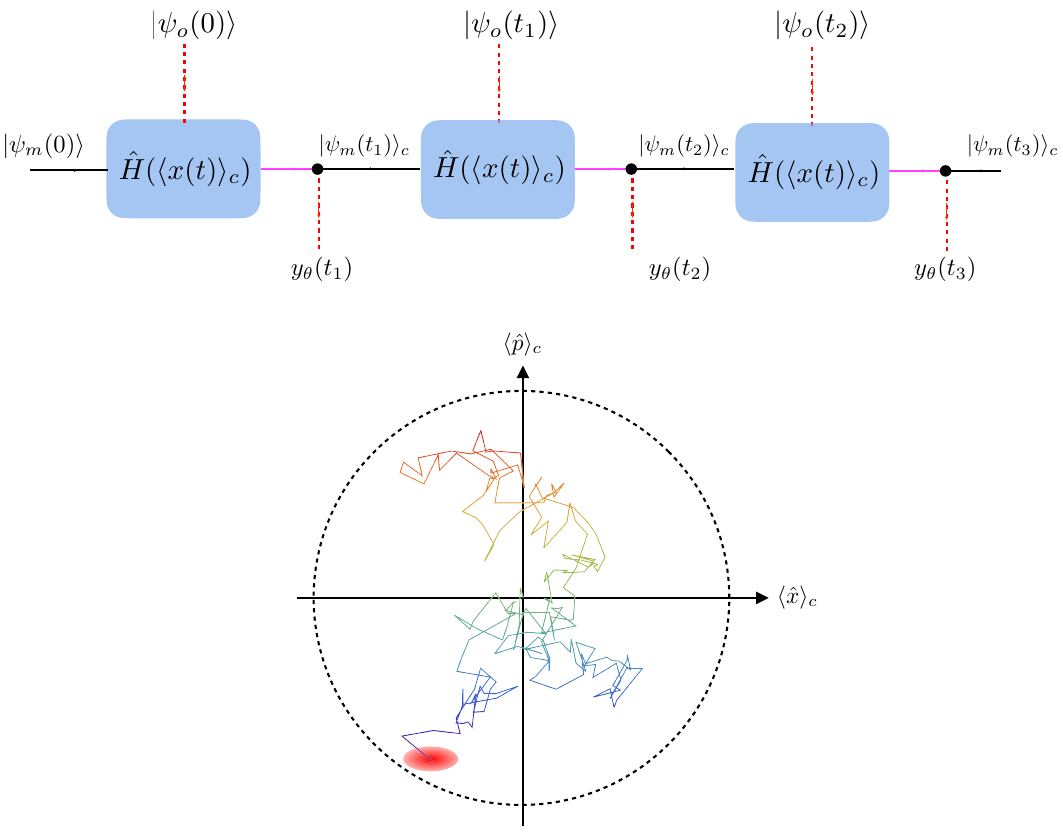}
\caption{Causal conditional evolution in the Schr{\"o}dinger picture. The optomechanical entangled state, created by pondermotive interaction, evolves under the SN nonlinear Hamiltonian. This state is projected into the mechanical Hilbert space and creates a conditional mechanical state that sources the self-gravity. The red, black, and magenta line describes the optical, mechanical, and optomechanical joint states, respectively. The lower panel illustrates the quantum trajectory, and the SN term will also fluctuates with the random evolution of the conditional mean mechanical displacement $\langle x\rangle_c$.}\label{fig:schemes}
\end{figure}

However, as discussed in\,\cite{Helou2017} and Paper I, this equivalency breaks down under the SN theory. The primary reason lies in the quantum evolution\,(or Hamiltonian) of the mechanical oscillator, which depends non-linearly on the quantum state. Specifically, within the \emph{pre/post-selection formalism}, the $\langle \hat x\rangle$ in Eqs.\,\eqref{eq:Hamiltonian_self_sn} and \eqref{eq:Hamiltonian_mutual_sn} represents the expectation value of $\hat x$ concerning the \emph{initial/final mechanical state}. In contrast, under the \emph{causal conditional formalism}, $\langle \hat x\rangle$ denotes the expectation value of $\hat x$ concerning the \emph{conditional mechanical state}. Consequently, the quantum trajectory generates classical gravity as the conditional mean of the mechanical displacement follows the quantum trajectory. Paper I investigated the causal conditional formalism within the Schr{\"o}dinger picture and established a stochastic master equation to describe the evolution of the mechanical state under SN theory, as briefly summarized below.

\subsection{Stochastic master equation and quantum trajectory}
For illustration, we first consider the self-gravity protocol. The stochastic master equation derived in Paper I is:
\be\label{eq:sme}
\begin{split}
d\hat\rho&=-\frac{i}{\hbar}[\hat{H}_0,\hat\rho]dt+\frac{\alpha}{\sqrt{2}}\sin\theta\{\hat{x}-\langle\hat{x}\rangle_c,\hat\rho\}dW\\&-\frac{i\alpha}{\sqrt{2}}\cos\theta[\hat{x},\hat\rho]dW-\frac{\alpha^2}{4}[\hat{x},[\hat{x},\hat\rho]]dt,
\end{split}
\ee
where $\hat\rho$ is the mirror's conditional density matrix, and $\hat H_0$ is the Hamiltonian from Eq.\,\eqref{eq:Hamiltonian_self_sn} with $\alpha=0$. The projective measurement on the outgoing light contributes the Lindblad and Ito terms. In this equation, the $\theta-$quadrature of the outgoing field $\hat a_\theta=\hat a_1\cos\theta+\hat a_2\sin\theta$ is measured, where $\hat a_1$ and $\hat a_2$ are amplitude and phase quadratures, respectively.

The quantum trajectory can be derived from this master equation:
\begin{equation}\label{eq:trajectory_self}
\begin{split}
d\langle\hat{x}\rangle_c=&\frac{\langle\hat{p}\rangle_c}{M}dt+\sqrt{2}\alpha\sin\theta V_{xx}dW,\\
d\langle\hat{p}\rangle_c=&-M\omega_{m}^2\langle x\rangle_c dt-\gamma_m\langle\hat{p}\rangle_c dt+\sqrt{2}\alpha\sin\theta V_{xp}dW\\
&+\frac{\hbar\alpha}{\sqrt{2}}\cos\theta dW,
\end{split}
\end{equation}
where $V_{xx}$ and $V_{xp}$ are second-order conditional moments satisfying Riccati equations:
\begin{equation}\label{eq:Ricatti_self}
\begin{split}
\dot{V}_{xx}=&\frac{2V_{xp}}{M}-2\alpha^2\sin^2\theta V_{xx}^2,\\
\dot{V}_{xp}=&-\gamma_m V_{xp}+\frac{V_{pp}}{M}-M\omega_{q}^2V_{xx}-2\alpha^2\sin^2\theta V_{xx}V_{xp}\\
&-\hbar\alpha^2\sin\theta\cos\theta V_{xx},\\
\dot{V}_{pp}=&-2\gamma_m V_{pp}-2M\omega_{q}^2V_{xp}-2\alpha^2\sin^2\theta V_{xp}^2+\frac{\hbar^2\alpha^2}{2}\\
&-2\hbar\alpha^2\sin\theta\cos\theta V_{xp}-\frac{\hbar^2\alpha^2}{2}\cos^2\theta.
\end{split}
\end{equation}
The measurement record of the output light is: 
\be\label{eq:measurement_self}
\tilde{y}_\theta=\alpha\sin\theta\langle \hat{x}\rangle+\frac{d W}{\sqrt{2}dt}.
\ee

To determine the statistical properties of the data series $y_\theta(t)$, integrate Eq.\,\eqref{eq:trajectory_self} to find the conditional mean $\langle\hat{x}(t)\rangle_c$ at the long-time tail:
 \begin{equation}\label{eq:xc_sch}
 \begin{split}
\langle\hat{x}(t)\rangle_c&=\sqrt{2}\alpha\int_0^tdW(s)e^{-\frac{\gamma_m }{2}(t-s)}\left(\frac{\hbar}{2M\tilde\omega_m}\sin{\tilde\omega_m(t-s)}\cos\theta\right.\\
&+V_{xx}(s)\left[\cos{\tilde\omega_m(t-s)}+\frac{\gamma_m}{2\tilde\omega_m}\sin{\tilde\omega_m(t-s)}\right]\sin\theta\\
&\left.+V^2_{xx}(s)\frac{\alpha^2}{\tilde\omega_m}\sin^3\theta\right),
 \end{split}
 \end{equation}

where the initial condition is ignored, and $\tilde\omega_m\equiv\sqrt{\omega_m^2-\gamma_m^2/4}$. With $\theta=\pi/2$ and using the approximation $\tilde\omega_m\approx \omega_m$, this formula reduces to the approximated case Eq.(11) in Paper I.

Using the Ito formula:
\begin{equation}
\begin{split}
\langle dW(t)dW(t')\rangle&=\frac{1}{T}\int_0^TdW(t+\tilde{t})dW(t'+\tilde{t})d\tilde{t}\\
&=dtdt'\delta(t-t'),
\end{split}
\end{equation}
and the measurement record formula Eq.\eqref{eq:measurement_self}, derive:
\begin{equation}\label{eq:spectrum_y_theta_sch}
\begin{split}
S_{y_\theta y_\theta}(\Omega)=\frac{|A^\theta_q(\Omega)-R_m(\Omega)|^2}{|R_m(\Omega)|^2},
\end{split}
\end{equation}
where 
\begin{equation}\label{eq:A_theta_q_R_m}
\begin{split}
&A^\theta_q(\Omega)=V_{xx}\alpha^2[V_{xx}\alpha^2\cos2\theta-V_{xx}\alpha^2-2(\gamma_m-i\Omega)]\sin^2\theta\\
&\qquad\quad-\alpha^2\sin\theta\cos\theta/M,\\
&R_m(\Omega)=\omega_m^2-\Omega^2-i\gamma_m\Omega.\\
\end{split}
\end{equation}

Assuming $\theta=\pi/2$, the spectrum simplifies to:
\begin{equation}
S_{\tilde{y}_{\pi/2}\tilde{y}_{\pi/2}}=\frac{4V_{xx}\alpha^2(V_{xx}\alpha^2+\gamma_m)(V_{xx}\alpha^2((V_{xx}\alpha^2+\gamma_m)+\omega_m^2)}{M^2|\omega_m^2-\Omega^2-i\gamma_m\Omega|^2},
\end{equation}
where $V_{xx}$ is the steady-state value determined by the Riccati equation, as given in Eq.\,\eqref{eq:Ricatti_self}. In the high-$Q$ limit, this result is consistent with those presented in Paper I.

\subsection{Revisit the solution of the master equation as a filtering problem}\label{sec:4.b}
The foregoing result can be re-derived from an alternative perspective, which will be valuable for comparison with the Heisenberg-picture approach discussed in the next section. Essentially, the quantum trajectory equations Eqs.\,\eqref{eq:trajectory_self}  have already contained the relationship between the measurement records and the conditional mean $(\langle x(t)\rangle_c,\langle p(t)\rangle_c)$. This relationship can be easily seen by representing the Ito term in quantum trajectory equations in terms of the measurement record (see Eq.\,\eqref{eq:measurement_self}):
\begin{widetext}
\begin{equation}\label{Expectation3}
\frac{d}{dt}
\left[
\begin{array}{c}
\langle\hat{x}\rangle_c\\
\langle\hat{p}\rangle_c
\end{array}
\right]
=
\left[
\begin{array}{cc}
-2\sin^2\theta\alpha^2V_{xx}&1/M\\
-M\omega_{m}^2-2\sin^2\theta\alpha^2V_{xp}-\hbar\alpha^2\sin\theta\cos\theta&-\gamma_m
\end{array}
\right]
\left[
\begin{array}{c}
\langle\hat{x}\rangle_c\\
\langle\hat{p}\rangle_c
\end{array}
\right]
+
\left[
\begin{array}{c}
2\sin\theta\alpha V_{xx}\\
2\sin\theta\alpha V_{xp}+\hbar\alpha\cos\theta
\end{array}
\right]\tilde{y}
\end{equation}
\end{widetext}
The solution of the above equation can be written as:
\be
\begin{split}
\langle\hat{x}(t)\rangle_c=&e^{-\Gamma t}\left[x_0\cos(\Omega_xt)+\frac{p_0}{M\omega_m}\sin(\Omega_xt)\right]\\
&+\int^t_{t_0}dt'K_\theta(t-t')\tilde{y}_\theta(t').
\end{split}
\ee
The filter function $K_\theta(t-t')$ is defined as:
\be\label{eq:filter_theta}
\begin{split}
K&_\theta(t-t')=2\alpha\sin\theta e^{-\Gamma (t-t')}V_{xx}(t')\cos\Omega_x(t-t')\\
&+2\alpha e^{-\Gamma (t-t')}\left[\frac{\cos\theta+MV_{xx}(t')\gamma_m\sin\theta}{2M\Omega_x}\sin\Omega_x(t-t')\right],
\end{split}
\ee
where
\be
\Gamma=\frac{\gamma_m+V_{xx}\alpha^2\sin^2\theta}{2},
\ee
and
\be
\begin{split}
\Omega^2_x=&\frac{(3V_{xx}^2\alpha^4+4V_{xx}\alpha^2\gamma_m-2\gamma_m^2+8\omega_m^2)+V_{xx}^2\alpha^4\cos(4\theta)}{8}\\
&+\frac{\alpha^2\sin(2\theta)/M-V_{xx}\alpha^2(V_{xx}\alpha^2+\gamma_m)\cos(2\theta)}{2}.\\
\end{split}
\ee
When we consider the long-term behavior\,($t\gg1/\Gamma$),
the initial condition will be forgotten and we have a mapping relation between $\langle \hat x(t)\rangle_c$ and $y_\theta(t')$ with the filter function $K_\theta(t-t')$ defined in Eq.\,\eqref{eq:filter_theta}, which is equivalent to Eq.\,\eqref{eq:xc_sch}. Moreover, in the above derivation, we used the approximation that the solutions of the Riccati equation will quickly converge to the steady state, i.e. $V_{xx}(t)$ is a constant:
\be
\begin{split}
&V_{xx}=\frac{-\gamma_m}{2\alpha^2\sin^2\theta}+\\
&\frac{\sqrt{\gamma_m^2-2\omega_q^2-\Lambda^2\sin2\theta+2\sqrt{\omega_q^4+\Lambda^4\sin^2\theta+\Lambda^2\omega_q^2\sin2\theta}}}{2\alpha^2\sin^2\theta}.
\end{split}
\ee
where $\Lambda=\alpha\sqrt{\hbar/M}$.

By applying the Fourier transformation to Eq.,\eqref{eq:filter_theta}, we derive the filter function in the frequency domain:
\be\label{eq:K_theta_AR_form}
\begin{split}
&\langle x(\Omega)\rangle_c=K_\theta(\Omega)\tilde y_\theta(\Omega)\\
&K_\theta(\Omega)=\frac{A^\theta_q(\Omega)}{\alpha\sin\theta(A^\theta_q(\Omega)-R_m(\Omega))}.
\end{split}
\ee
When combined with the Fourier form of Eq.,\eqref{eq:measurement_self}:
\be
\tilde y_{\theta}(\Omega)=\alpha\sin\theta \langle x(\Omega)\rangle_c+dW(\Omega)/\sqrt{2},
\ee
one can easily rederive Eq.,\eqref{eq:spectrum_y_theta_sch}.

Similar findings for the mutual gravity protocol are discussed in Paper I. However, a notable limitation of the stochastic master equation approach is its inability to investigate the entanglement structure of the outgoing lights within the mutual gravity protocol.

\subsection{Limitations of stochastic master equation}
The quantum entanglement of a bipartite Gaussian state can be characterized by the logarithmic negativity $\epsilon_N$, which is defined as:
\begin{equation}\label{eq:logarithmic_negativity}
\varepsilon_N={\rm max}\{-(1/2){\rm ln}\left[\left(\Sigma-\sqrt{\Sigma^2-4{\rm det}\sigma}\right)/2\right],0\}
\end{equation}
with $\Sigma={\rm det}\sigma_A+{\rm det}\sigma_B-2{\rm det}\sigma_{AB}$,
\begin{equation}\label{eq:covariance_matrix_general}
\sigma_A=
\left[
\begin{array}{cc}
S_{\tilde{y}_{1A}\tilde{y}_{1A}}&S_{\tilde{y}_{1A}\tilde{y}_{2A}}\\
S_{\tilde{y}_{2A}\tilde{y}_{1A}}&S_{\tilde{y}_{2A}\tilde{y}_{2A}}
\end{array}
\right],
\quad
\sigma_{AB}=
\left[
\begin{array}{cc}
S_{\tilde{y}_{1A}\tilde{y}_{1B}}&S_{\tilde{y}_{1A}\tilde{y}_{2B}}\\
S_{\tilde{y}_{2A}\tilde{y}_{1B}}&S_{\tilde{y}_{2A}\tilde{y}_{2B}}
\end{array}
\right],
\end{equation}
where $\tilde{y}_{1A/B},\tilde{y}_{2A/B}$ are the recorded data strings when measuring the amplitude and phase quadratures of the outgoing light fields. To obtain the logarithmic negativity $\epsilon_N$, the covariance matrices $\sigma_A$ and $\sigma_{AB}$ are required\,\cite{Simon2000,Duan2000}.

However, computing cross-correlation terms such as $S_{\tilde{y}_{1A}\tilde{y}_{2A}}$ via the stochastic master equation approach is not feasible. This limitation arises because the stochastic master equation is derived for a fixed measurement quadrature angle $\theta$, and the ensemble average with respect to the Ito term $dW$ can only be performed for $\tilde y_\theta$. Calculating the correlation between $\tilde y_\theta$ and $\tilde y_{\theta'}$ are not meaningful within this framework.

This problem also exists in standard quantum mechanics and quantum gravity cases; however, it can be circumvented using the Heisenberg-picture approach\,(e.g. the method used in\,\cite{Miao2020}). Therefore, to compute the cross-correlation function in the SN theory, a corresponding Heisenberg picture approach is necessary.

\section{Causal conditional Heisenberg dynamics as a Wiener filtering problem}\label{sec:generation_functional}
The Heisenberg picture provides a framework for examining the dynamical evolution of quantum operators. In this section, we present a transformation from the Schrödinger picture to the Heisenberg picture within the context of SN-type nonlinear quantum mechanics. This transformation characterizes the evolution of the conditional quantum state as an optimal filtering problem.

In the causal conditional formalism, the conditional density matrix of the mechanical oscillator is expressed as\,\cite{Khalili2010,Ebhardt2009}:
\be
\hat\rho_c=\langle y_N|\hat U_{N-1}...\hat U_{1}\langle y_1|\hat U_0\hat \rho_0 \hat U^\dag_0 | y_1\rangle\hat U^\dag_{1}...\hat U^\dag_{N-1}|y_N\rangle,
\ee
where $|y_i\rangle$ is the eigenvector of the outgoing light field and $y_i$ is the optical phase measurement data. The $\hat U_i=\hat U(t_i,\langle x(t_i)\rangle_c)$ is the evolution operator of the time interval $[t_i, t_i+\Delta t]$ which depends on the conditional mean displacement, a hallmark of SN theory under casual conditional prescription. This density matrix can alternately be represented as:
\be
\hat\rho_c={\rm Tr}_o[|y_N\rangle\langle y_N|\otimes...\otimes|y_1\rangle\langle y_1|\hat U_{N-1}...\hat U_0\rho_0\hat U^\dag_0...\hat U^\dag_{N-1}],
\ee
and in the continuous limit we have:
\be
\hat\rho_c={\rm Tr}_o[\mathcal{\hat P}\mathcal{\hat U}\rho_0\mathcal{\hat U}^\dag]={\rm Tr}_o[\mathcal{\hat U}\rho_0\mathcal{\hat U}^\dag\mathcal{\hat P}],
\ee
where
\be\label{eq:discretize_to_continuum}
\begin{split}
&\Pi^N_{i=1}\delta(\hat y_i-y_i)\rightarrow\int D[k(t')]{\rm exp}\left[i\int^t_{-\infty}dt'k(t')(\hat y(t)-y(t))\right],\\
&\hat U_{N-1}...\hat U_0\rightarrow\mathcal{\hat U}(t,0)=\mathcal{T}\left[{\rm exp}[-i\int^t_0dt'\hat H(\langle x(t')\rangle_c)]\right].
\end{split}
\ee
The generating function is given by:
\be
J(\alpha)={\rm Tr}_m[\hat\rho_c(t)e^{i\alpha \hat x}]={\rm Tr}[\mathcal{\hat U}\rho_0\mathcal{\hat U}^\dag\mathcal{\hat P}e^{i\alpha \hat x}].
\ee
Utilizing the permutation symmetry in the trace operation and $\mathcal{\hat U}\mathcal{\hat U}^\dag=1$, we find:
\be
J(\alpha)={\rm Tr}[\hat\rho_0\mathcal{\hat U}^\dag\mathcal{\hat P}\mathcal{\hat U}\mathcal{\hat U}^\dag e^{i\alpha \hat x}\mathcal{\hat U}]={\rm Tr}[\hat\rho_0\mathcal{\hat P}^H e^{i\alpha \hat x^H}],
\ee
where the Heisenberg picture operators are defined as:
\be\label{eq:Heisenberg_continuum}
\begin{split}
e^{i\alpha \hat x^H(t)}&=\mathcal{\hat U}^\dag(t,0) e^{i\alpha \hat x}\mathcal{\hat U}(t,0), \\
\mathcal{P}^H&=\mathcal{\hat U^\dag}(t,0)\mathcal{\hat P}\mathcal{\hat U}(t,0)\\
&=\int D[k(t')]{\rm exp}\left[i\int^t_{-\infty}dt'k(t')(\hat y^H(t)-y(t))\right]
\end{split}
\ee
This allows the generating function to be represented using Heisenberg picture operators, solvable via Heisenberg equations of motion in SN theory.

For a general theory with the classical-quantum mixture feature that\,(SN theory has such a feature as we shall see later):
\be
\begin{split}
&\hat y^H(t)=\hat y^H_q(t)+y^H_{\rm cl}(t),\\
&\hat x^H(t)=\hat x^H_q(t)+x^H_{\rm cl}(t),
\end{split}
\ee
the generating functional is further formulated as:
\be
\begin{split}
J(&\alpha)={\rm exp}\left[i\alpha x^H_{\rm cl}(t)+i\int^t_{-\infty}dt'k(t')[y^H_{\rm cl}(t')-y(t')]\right]\\
&\int D[k(t')]{\rm Tr}\left[\hat\rho_0{\rm exp}\left(i\alpha\hat x^H_q(t)+i\int^t_{-\infty}dt'k(t')y_q^H(t')\right)\right].
\end{split}
\ee
The conditional density matrix thus becomes:
\be
\begin{split}
&\hat\rho_c={\rm Tr}_o[\mathcal{\hat U}\rho_0\mathcal{\hat U}^\dag\mathcal{\hat P}]={\rm Tr}_o[\rho_0\mathcal{\hat U}^\dag\mathcal{\hat P}\mathcal{\hat U}]={\rm Tr}_o[\rho_0\mathcal{\hat P}^H]\\
&={\rm Tr}_o\left[\rho_0\int D[k(t')]{\rm exp}\left[i\int^t_{-\infty}dt'k(t')(\hat y^H(t)-y(t))\right]\right],
\end{split}
\ee
and for a Gaussian process:
\be
\begin{split}
&{\rm Tr}\left[\hat\rho_0{\rm exp}\left(i\alpha\hat x^H_q(t)+i\int^t_{-\infty}dt'k(t')y_q^H(t')\right)\right]\\
&={\rm exp}\left[-\frac{1}{2}\left\langle \left[\alpha\hat x^H_q(t)+\int^t_{-\infty}dt'k(t')y_q^H(t')\right]^2\right\rangle\right],
\end{split}
\ee
where $\langle...\rangle={\rm Tr}[\rho_0...]$.

Expressing
\be
\hat x^H_q(t)=\int^t_{-\infty}K_q(t-t')\hat y_q^H(t')dt'+\hat R(t),
\ee
with $\langle \hat y_q^H(t')\hat R(t)\rangle=0, \forall t'<t$, meaning \(K_q(t-t')\) is the optimal estimator of \(\hat x_q(t)\) from the data string \(y_q^H(t')\), results in:
\be
\begin{split}
&-\frac{1}{2}\left\langle \left[\alpha\hat x^H_q(t)+\int^t_{-\infty}dt'k(t')y_q^H(t')\right]^2\right\rangle=\\
&-\frac{1}{2}\left[\alpha\langle\hat R(t)^2\rangle+\int\int dt'dt''\tilde K(t,t')\tilde K(t,t'')\langle \hat y_q^H(t')\hat y_q^H(t'')\rangle\right],
\end{split}
\ee
where the cross term $\propto \langle\hat y_q^H(t')\hat R(t)\rangle=0$, and $\tilde K(t,t')$ is defined as:
\be
\tilde K(t,t')=\alpha K_q(t-t')+k(t').
\ee

Substituting the above equation into the generating functional yields:
\be
\begin{split}
J(\alpha)=&{\rm exp}\left[i\alpha x^H_{\rm cl}(t)-\frac{1}{2}\alpha^2\langle\hat R(t)^2\rangle\right]\int D[k(t')]\\
&{\rm exp}\left[i\int^t_{-\infty}dt'[\tilde{K}(t,t')-\alpha K_q(t-t')][y^H_{\rm cl}(t')-y(t')]\right]\\
&{\rm exp}\left[-\frac{1}{2}\int\int dt'dt''\tilde K(t,t')\tilde K(t,t'')\langle \hat y_q^H(t')\hat y_q^H(t'')\rangle\right].
\end{split}
\ee
Changing the functional variable:
\be
\int D[k(t')]\rightarrow\int D[\tilde K(t,t')],
\ee
leads to
\be\label{eq:functional_ktilde}
\begin{split}
&J(\alpha)=\\
&{\rm exp}\left[i\alpha x^H_{\rm cl}(t)-\frac{1}{2}\alpha^2\langle\hat R(t)^2\rangle-i\alpha\int^t_{\infty}K_q(t-t')[y^H_{\rm cl}(t')-y(t')]\right]\\
&\int D[\tilde K(t,t')]\left\langle{\rm exp}\left[i\int^t_{-\infty}dt'\tilde{K}(t,t')[\hat y^H(t')-y(t')]\right]\right\rangle,
\end{split}
\ee
where the the following two equalities are used:
\be
\begin{split}
&\left\langle{\rm exp}\left[i\int \tilde K(t,t')\hat y_q^H(t')dt'\right]\right\rangle=\\
&{\rm exp}\left[-\frac{1}{2}\int\int dt'dt''\tilde K(t,t')\tilde K(t,t'')\langle \hat y_q^H(t')\hat y_q^H(t'')\rangle\right],\\
&\hat y^H(t')=\hat y^H_q(t')+\hat y^H_{\rm cl}(t').
\end{split}
\ee
The functional integral in Eq.\,\eqref{eq:functional_ktilde} is unity since it represents ${\rm Tr}_m[\rho_c]=1$.

Finally, the generation functional is:
\be
\begin{split}
J(\alpha)=&{\rm exp}\left[-\frac{1}{2}\alpha^2\langle\hat R(t)^2\rangle\right]\\
&{\rm exp}\left[i\alpha x^H_{\rm cl}(t)-i\alpha\int^t_{-\infty}K_q(t-t')[y^H_{\rm cl}(t')-y(t')]\right]
\end{split}
\ee
and the conditional mean is:
\be\label{eq:conditional_mean_generation_functional}
\langle \hat x(t)\rangle_c=\left.\frac{\partial}{\partial \alpha}J(\alpha)\right|_{\alpha=0}
=x^H_{\rm cl}(t)+\int^t_{-\infty}K_q(t-t')y^H_q(t'),
\ee
where $y^H_q(t')\equiv y(t')-y^H_{\rm cl}(t')$.
The filter function $K_q(t-t')$ can be determined using the Wiener-Hopf method, familiar with standard quantum mechanics. This involves solving the Wiener-Hopf equation, $\langle\hat y_q^H(t')\hat R(t)\rangle=0$:
\be\label{eq:WHequation}
C_{\hat x_q^H\hat y_q^H}(t,t')=\int^t_{-\infty}K_q(t-t')C_{\hat y_q^H\hat y_q^H}(t,t'),\,\,\forall t'<t,
\ee
where $C_{\hat A\hat B}$ denotes the correlation function.
As we will explore in the next section, within the SN theory, mechanical dynamics are partly driven by the conditional mean displacement $\langle x^H\rangle_c$ via classical gravity\,\cite{Helou2017}. The Wiener filter method developed here allows for the calculation of $\langle x^H\rangle_c$, which is crucial for determining expressions for $\hat x^H(t)$ and the outgoing field quadrature data as influenced by mirror displacement. Given the relationship between the amplitude and phase quadratures of the incoming optical field and outgoing optical field quadrature at a homodyne detection angle $\theta$ in the Heisenberg picture, along with the known correlation of the amplitude and phase quadratures of the incoming field:
\be
\langle \hat a_i(t)\hat a_j(t')\rangle_{\rm sym}=\delta_{ij}\delta (t-t'),
\ee
where $\langle A(t)\hat B(t')\rangle_{\rm sym}={\rm Tr}[\hat \rho(\langle A(t)\hat B(t')+\langle B(t')\hat A(t))]/2$ defines the correlation function for two Hermitian operators $\hat A,\hat B$, and $(i,j=1,2)$ denote the amplitude and phase quadratures, it becomes straightforward to compute the correlation spectrum for any two outgoing optical field quadratures. This computation subsequently facilitates the calculation of logarithmic negativity.

\section{Heisenberg-picture approach to the optomechanical protocols}\label{sec3}
In the previous section, we can see that one key step in obtaining the filter function $K_q(t-t')$ is to compute the correlation function of $\hat x_q^H,\hat y_q^H$, which means we need to find the expression of the displacement/field quadrature decomposition $\hat x^H=\hat x^H_q+ x^H_{\rm cl}, \hat y^H=\hat y^H_q+y^H_{\rm cl}$. For brevity, we will omit all superscripts $H$ of the operators in the following text, since we are working in the Heisenberg picture.  The expressions of $\hat x(t),\hat y(t)$ can be obtained by solving the Heisenberg equations of motion for the mechanical oscillator in the SN theory:
\begin{equation}\label{eq:eom_self_mech}
\begin{split}
&\dot{\hat{x}}=\frac{\hat{p}}{M},\\
&\dot{\hat{p}}=-M(\omega_m^2+\omega^2_{SN})\hat{x}+M\omega^2_{SN}\langle\hat{x}\rangle_c+\hbar\alpha\hat{a}_1.\\\\
\end{split}
\end{equation}
Here, the SN gravity potential affects the test mass in two ways: (1) it shifts the resonant frequency of $\hat x$ from $\omega_m$ to $\omega_q=\sqrt{\omega_m^2+\omega_{SN}^2}$; (2) within the causal-conditional formalism, there is an interpretation where a "spectator," aware of all measurement records, applies the self-gravity force $M\omega^2_{SN}\langle\hat{x}\rangle_c$ on the test mass.

The input-output relation for the optical field is:
\be\label{eq:in_out_self}
\hat{b}_\theta=\cos\theta\hat{a}_1+\sin\theta\hat{a}_2+\sin\theta\alpha\hat{x}.
\ee
The solution of Eq.\eqref{eq:eom_self_mech} is:
\begin{equation}
\hat{x}(t)=\int_{-\infty}^tdt'\chi_q(t-t')[\alpha\hat{a}_1(t')+M\omega_{SN}^2\langle\hat{x}(t')\rangle_c],
\end{equation}
where $\chi_q$ is the quantum mechanical response function with Fourier-transformed form as 
\be
\chi_q(\omega)=\frac{1}{m(-\omega^2+\omega_q^2-i\gamma_m\omega)}.
\ee
This $\hat{x}(t)$ can be decomposed as
\begin{equation}
\begin{split}
&\hat{x}(t)=\hat{x}_q(t)+x_{\rm cl}(t),\\
&\hat{x}_q(t)=\int_{-\infty}^tdt'\chi_q(t-t')\alpha\hat{a}_1(t'),\\
&x_{\rm cl}(t)=\int_{-\infty}^tdt'\chi_q(t-t')M\omega_{SN}^2\langle\hat{x}(t')\rangle_c.
\end{split}
\end{equation}
The $\hat x_q(t)$ is the displacement operator driven by the environmental quantum noise\,(e.g. radiation pressure noise), and $x_{\rm cl}(t)$ is a classical displacement driven by the classical gravity sourced by conditional mean displacement $\langle\hat{x}(t')\rangle_c$ for $t'<t$.

Substituting the solution of Eq.\eqref{eq:eom_self_mech} into the above input-output relation leads to:
\be\label{eq:in_out_sn_self}
\begin{split}
\hat{b}_{\theta}(t)=&\cos\theta\hat{a}_1+\sin\theta\hat{a}_2+\sin\theta\alpha\int_{-\infty}^t\chi_q(t-t')\hbar\hat{a}_1(t')dt'\\
&+\sin\theta\alpha \int_{-\infty}^t\chi_q(t-t')M\omega_{SN}^2\langle\hat{x}(t')\rangle_c dt',
\end{split}
\ee
where the field quadrature with homodyne angle $\theta$ is measured, and this $\hat b_\theta$ corresponds to the $\hat y_q^H(t)$ in the previous section. Measuring the outgoing field $\hat b_\theta(t)$ yields a data string $\tilde y_\theta$\,(which indicates the commutation relation $[\hat{b}_{\theta}(t),\hat{b}_{\theta}(t')]=0$). This data string $\tilde y_\theta$ comprises two components: the term from the classical conditional mean mechanical displacement $\langle\hat{x}(t)\rangle_c$, and the quantum term $y^q_\theta(t)$:
\be\label{eq:in_out_sn_self_quan}
\hat b_\theta^q(t)=\hat b_\theta(t)-\sin\theta\alpha \int_{-\infty}^t\chi_q(t-t')M\omega_{SN}^2\langle\hat{x}(t')\rangle_c dt',
\ee
which also satisfies $[\hat{b}^q_{\theta}(t),\hat{b}^q_{\theta}(t')]=0$.

\subsection{Causal-conditional dynamics as a Wiener filtering problem}
Following the discussion in Section\,\ref{sec:generation_functional}, we now analyse the causal-conditional dynamics in SN theory as a Wiener filtering problem\,\cite{Bouten2007,wiener1949extrapolation,Ebhardt2009}.

\begin{widetext}
\begin{figure*}
\centering
\includegraphics[width=0.85\textwidth]{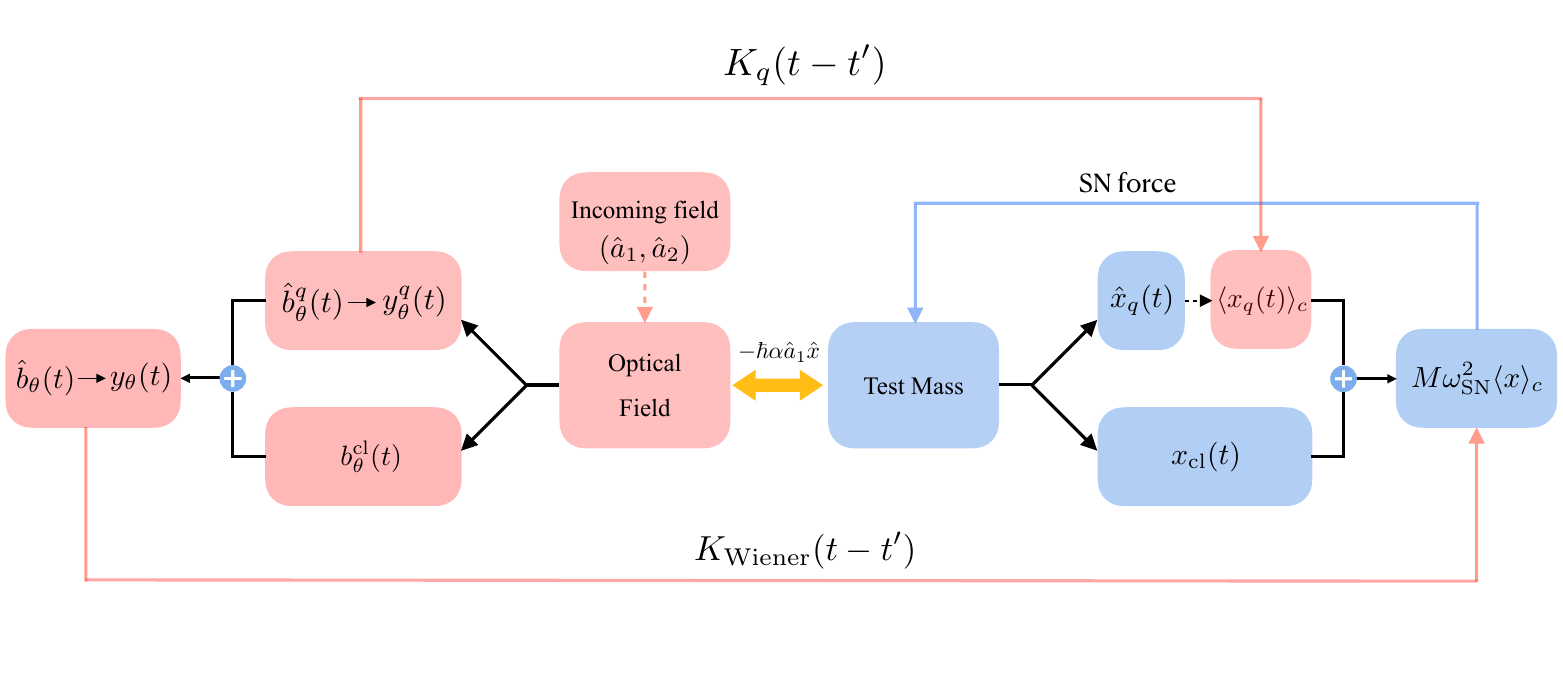}
\caption{In the Heisenberg picture, causal conditional dynamics can be approached as a Wiener filtering problem, particularly within the self-gravity protocol of the SN theory. Here, the conditional displacement of the test mass, $\langle x(t)\rangle_c$, is composed of two distinct components. The first component arises from the quantum radiation pressure force and can be determined by filtering the quantum component of the outgoing field $\hat b_q^\theta$. The second component is a result of the classical gravity force, which is influenced by the quantum trajectory $\langle x(t')\rangle_c$ at earlier times $t<t'$. The accompanying flowchart outlines the method for calculating these causal conditional dynamics, as detailed in the main text. Note that the $\langle x_q(t)\rangle_c$ is obtained through Wiener-filtering the optical data, which provides an optimal estimation of the $\hat x_q(t)$. This is why the $\langle x_q(t)\rangle_c$ is represented in pink.}
\label{fig:heisenberg_wiener}
\end{figure*}
\end{widetext}

Substituting the formulae of $\hat b^q_\theta(t),\hat x_q(t)$ into the Wiener-Hopf equation\,\eqref{eq:WHequation}, we can obtain the Wiener filter function $K_q(\Omega)$ in the frequency domain using the famous Wiener-Hopf method (for a detailed derivation, see Appendix):
\begin{equation}\label{eq:Wiener_filter}
\begin{split}
K_{q}(\Omega)
&=\frac{\beta\beta_c+\Omega(\beta-\beta_c)+i\gamma_m\Omega-\omega_m^2}{\alpha\sin\theta(\beta-\Omega)(\Omega+\beta_c)}\\
&=\frac{A^\theta_q(\Omega)}{\alpha\sin\theta(A^\theta_q(\Omega)-R_q(\Omega))},
\end{split}
\end{equation}
where $R_q(\Omega)=\omega_q^2-\Omega^2-i\gamma_m\Omega$ and $A^\theta_q(\Omega)$ has the same form as Eq.\,\eqref{eq:K_theta_AR_form}. In the first line, $\beta=\beta^*_c$ are the roots of the numerator of the spectrum $S_{y^q_\theta y^q_\theta}$. Detailed expressions for these roots are provided in Appendix\,\ref{subsec:Wiener_filter}\,(note that the $\omega_m$ in Appendix\,\ref{subsec:Wiener_filter} should be replaced by $\omega_q$ in the SN theory). The validity of the second equality is demonstrated in Appendix\,\ref{subsec:filter_w}.

Using Eq.\,\eqref{eq:conditional_mean_generation_functional}:
\be
\begin{split}
\langle x(t)\rangle_c=&\int_{-\infty}^tdt' K_{q}(t-t')\tilde{y}_\theta^q(t')\\
&+\int_{-\infty}^tdt'\chi_q(t-t')M\omega_{SN}^2\langle\hat{x}(t')\rangle_c,
\end{split}
\ee 
and applying a Fourier transformation
leads to the conditional mean displacement in the frequency domain $\langle x(\Omega)\rangle_c$:
\begin{equation}\label{eq:x_y}
\langle\hat{x}(\Omega)\rangle_c=\frac{\chi_m(\Omega)}{\chi_q(\Omega)}K_{q}(\Omega)\tilde{y}^q_\theta(\Omega).
\end{equation}
When overviewing the Schr{\"o}dinger-picture approach, we have derived the filter function using an alternative method based on the master equation\,\eqref{eq:trajectory_self}. To facilitate a comparison between these two approaches, we require another form of the filter function that relates the conditional mean displacement $\langle \hat x(\Omega)\rangle_c$ to the complete data string $\tilde y_\theta(\Omega)$. By representing the $\hat y_q(\Omega)$ in terms of $\langle \hat x(\Omega)\rangle_c$ using Eq.\,\eqref{eq:x_y} and substituting into:
\be\label{eq:y_theta_omega}
\tilde y_\theta(\Omega)=\tilde y^q_\theta(\Omega)+\sin\theta\alpha\chi_q(\Omega)M\omega_{\rm SN}^2\langle x(\Omega)\rangle_c,
\ee
leads to a filter function $K_{\rm Wiener}(\Omega)$:
\be\label{eq:Wiener_filter_sn}
\begin{split}
&\langle\hat{x}(\Omega)\rangle_c=K_{\rm Wiener}(\Omega)\tilde{y}_\theta(\Omega),\\
&K_{\rm Wiener}(\Omega)=\frac{A^\theta_q(\Omega)}{\alpha\sin\theta(A^\theta_q(\Omega)-R_m(\Omega))},
\end{split}
\ee
where $A^\theta_q(\Omega)$ and $R_m(\Omega)$ are given in Eqs.\,\eqref{eq:A_theta_q_R_m}. This filter function is the same as the $K_\theta(\Omega)$\,(Eq.\eqref{eq:K_theta_AR_form}) derived in the Schr{\"o}dinger picture.

Complementary, if we directly substitute Eq.\,\eqref{eq:x_y} into Eq.\,\eqref{eq:y_theta_omega}, we can rewrite Eq.\,\eqref{eq:y_theta_omega} in terms of $\tilde{y}^q_{\theta}$ and finally compute the covariance matrix of the outgoing light field which will be discussed in Section\,\ref{sec:4.c}.

\subsection{Spectrum and covariance matrix for the outgoing field}\label{sec:4.c}
Utilizing the filter function to compute the conditional mean motion in the SN theory enables the calculation of the covariance matrix for the outgoing field.

The measurement record $\tilde y_\theta$ can be expressed as:
\be
\tilde y_\theta(\Omega)=\tilde y^q_\theta(\Omega)
+M\omega_{\rm SN}^2\sin\theta\alpha\chi_q(\Omega)K_{\rm Wiener}(\Omega)\tilde y_\theta(\Omega).
\ee
By substituting the Wiener filter $K_{\rm Wiener}(\Omega)$ into Eq.\eqref{eq:Wiener_filter_sn}, we obtain:
\be\label{eq:yq-y_relation_self}
\begin{split}
&\tilde y_\theta(\Omega)=K^\theta_{yy}(\Omega)\tilde y^q_\theta(\Omega),\\
&\text{where }K^\theta_{yy}(\Omega)=\frac{R_q(\Omega)}{R_m(\Omega)}\frac{A^\theta_q(\Omega)-R_m(\Omega)}{A^\theta_q(\Omega)-R_q(\Omega)}.
\end{split}
\ee

The noise spectrum of the outgoing field is then computed as:
\be
S_{\tilde y_\theta\tilde y_\theta}(\Omega)=\left|K^\theta_{yy}(\Omega)\right|^2S_{\tilde y^q_\theta\tilde y^q_\theta}(\Omega),
\ee
where $S_{\tilde y^q_\theta\tilde y^q_\theta}(\Omega)$ is given by the standard optomechanical system:
\be
S_{\tilde y^q_\theta\tilde y^q_\theta}(\Omega)=
1+\alpha^2\sin2\theta{\rm Re}\left[\chi_q(\Omega)\right]+\alpha^4\sin^2\theta|\chi_q(\Omega)|^2.
\ee
Appendix B demonstrates that $S_{\tilde y^q_\theta\tilde y^q_\theta}(\Omega)$ satisfies the equality:
\be
S_{\tilde y^q_\theta\tilde y^q_\theta}(\Omega)=\left|\frac{A^\theta_q(\Omega)-R_q(\Omega)}{R_q(\Omega)}\right|^2.
\ee
Thus, we deduce:
\be
S_{\tilde y_\theta\tilde y_\theta}(\Omega)=\left|\frac{A^\theta_q(\Omega)-R_m(\Omega)}{R_m(\Omega)}\right|^2,
\ee
which is exactly equivalent to the result we obtained using the Schr{\"o}dinger picture in Eq.\,\eqref{eq:spectrum_y_theta_sch} in Section\,\ref{sec2},.

In addition, the environmental effect needs to be considered in the calculation of the spectrum and covariance matrix.  Within the framework of nonlinear quantum mechanical theories, there are two different prescriptions to treat the thermal environment, and this subtlety was extensively discussed in\,\cite{Helou2017}. Generally speaking, these two prescriptions correspond to treating the thermal noise classically or quantum-mechanically. In the main text, we focus on quantum thermal noise, while classical thermal noise is discussed in detail in the Appendix.\,\ref{sec:classical_thermal}. Quantum thermal noise in this context includes all unmonitored environmental noise sources that lead to decoherence of the quantum state. This interpretation treats environmental degrees of freedom as part of the quantum mechanical system. The quantum thermal noise drives the quantum motion of the test mass and affects the outgoing optical quadrature as,
\be
\begin{split}
\hat{b}^q_{\theta}(t)=&\cos\theta\hat{a}_1+\sin\theta\hat{a}_2+\sin\theta\alpha\int_{-\infty}^t\chi_q(t-t')\hbar\hat{a}_1(t')dt'\\
&+\sin\theta\alpha \int_{-\infty}^t\chi_q(t-t')\hat{F}_{\rm th} dt',
\end{split}
\ee

The thermal noise spectrum approximation at high-temperature is:
\begin{equation}
\begin{split}
S_{F_{\rm th}F_{\rm th}}(\Omega)&=4\hbar\left[\frac{1}{e^{\frac{\hbar\Omega}{k_BT}}-1}+\frac{1}{2}\right]\frac{{\rm Im}[\chi_m(\Omega)]}{|\chi_m(\Omega)|^2}\\
&\approx4k_BTM\gamma_m+2\hbar\Omega M\gamma_m.
\end{split}
\end{equation}
The quantum thermal noise modifies the Wiener filter function, in which the factor $a_2$ of the $\beta$ (details in Appendix A.1) is modified as:
\begin{equation}\label{eq:beta_thermal}
a_2=2\tilde{\Lambda}^4(1-\cos2\theta)+4\omega_q^2(\omega_q^2+\Lambda^2\sin2\theta)
\end{equation}
where $\tilde\Lambda^2=\Lambda^2+4\gamma_mk_BT/\hbar$.

Since the mathematical structure of the above formulas is unchanged by the thermal noise, the filter function $K_q(\Omega)$ is also formally unchanged. Then it is easy to show that the $S_{\tilde y_\theta\tilde y_\theta}$ also takes the same form, except that the $\beta$ substituted in $A_q^\theta(\Omega)$ (seen Appendix.B) should be given by the above formula Eq.\eqref{eq:beta_thermal}. Straightforward algebra shows that the outgoing field with the contribution of a quantum thermal bath is:
\begin{equation}
\begin{split}
S_{\tilde{y}_\theta\tilde{y}_\theta}=&1+|\chi_m|^2\alpha^4\sin^4\theta+|\chi_m|^2\alpha^2\sin2\theta(\omega_m^2-\Omega^2)\\
&+4|\chi_m|^2\sin\theta^2\alpha^2M\gamma_mk_BT\\
&+\omega_{\rm SN}^2M^2|\chi_m|^2\left[2\omega_q^2+\Lambda^2\sin2\theta\right.\\
&\left.-2\sqrt{\Lambda^2\tilde\Lambda^2\sin^2\theta+\omega_q^2(\omega_q^2+2\Lambda^2\sin2\theta)}\right].
\end{split}
\end{equation}

The components of this spectrum have transparent physical meaning: the first line accounts for quantum shot noise and back-action noise\,\cite{Chen_2013}, the second line arises from the thermal bath's direct impact on $\hat x-$dynamics, and the final two lines describe the effect of classical self-gravity\,(SN term) modified by the quantum thermal bath. Basically, the evolution of the $\langle \hat x(t)\rangle_c$ depends on $V_{xx}$, which is affected by the thermal bath. Consequently, the self gravity force $-M\omega^2_{\rm SN}\langle \hat x(t)\rangle_c$, which drives the test mass, is also affected by the thermal bath. The spectrum $S_{\tilde{y}_\theta\tilde{y}_\theta}$ can also be derived using the Wiener filter method.

\begin{table}[h!]
    \centering
    \begin{tabular}{|c|c|c|}
    \hline
Parameters&Symbol&Value\\
\hline
Mirror mass&$M$&0.2\,kg\\
\hline
Mirror bare frequency&$\omega_m/(2\pi)$&$4\times10^{-3}\,{\rm Hz}$\\
\hline
SN frequency&$\omega_{\rm SN}/(2\pi)$&$7.8\times10^{-2}\,{\rm Hz}$\\
\hline
Quality factor&$Q_m$&$10^7$\\
\hline
Mechanical damping&$\gamma_m/(2\pi)$&$4\times10^{-10}\,{\rm Hz}$\\
\hline
Optical wavelength&$\lambda$&$1064\,{\rm nm}$\\
\hline
Cavity Finesse&$\mathcal{F}$&$300$\\
\hline
Intra-cavity power&$P_{\rm cav}$&$480\,{\rm nW}$\\
\hline
    \end{tabular}
    \caption{Sample parameters for the self-gravity optomechanical protocol.}
    \label{tab:self_gravity}
\end{table}

\begin{widetext}
\begin{figure*}
\centering
\includegraphics[width=0.85\textwidth]{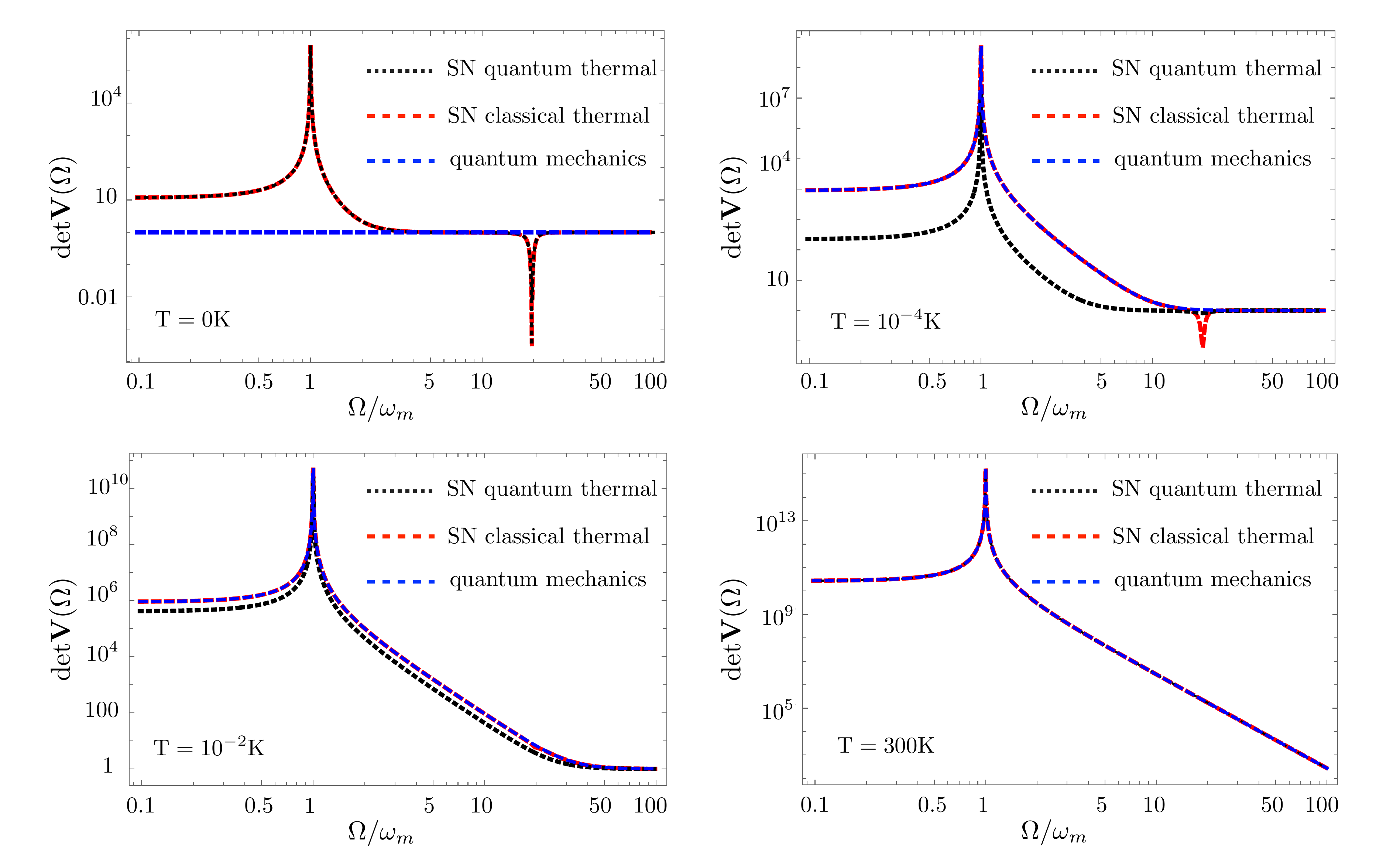}
\caption{The ${\rm det}\mathbf{V}(\Omega)$ at different temperatures of SN theory and quantum gravity for the self-gravity protocol. Note that the ${\rm det}\mathbf{V}(\Omega)$ can be smaller than one in the SN case when the temperature is sufficiently low. When $T=0$, the dip of ${\rm det}\mathbf{V}(\Omega)$  in the SN case locates at $\Omega=\omega_q$, which originates from the factor $K^{\pi/2}_{yy}(\Omega)$. We plot three different cases, the black and red curves are the ${\rm det}\mathbf{V}(\Omega)$ in the SN theory where the thermal noise is treated quantum mechanically and classically, while the blue curve is the ${\rm det}\mathbf{V}(\Omega)$ in standard quantum mechanics. The details of classical thermal noise are discussed in Appendix.\,\ref{sec:classical_thermal}}\label{fig:SN_detV}
\end{figure*}
\end{widetext}

Considering the thermal bath, we can also compute the cross-correlation function using Eq\,\eqref{eq:yq-y_relation_self}. The resulting covariance matrix is:
\begin{equation}
\begin{split}
\mathbf{V}(\Omega)=\left[
\begin{array}{cc}
S_{\tilde{y}_0^q\tilde{y}_0^q}(\Omega)&K^{\pi/2*}_{yy}(\Omega)S_{\tilde{y}_0^q\tilde{y}_{\pi/2}^q}(\Omega)\\
K^{\pi/2}_{yy}(\Omega)S_{\tilde{y}_{\pi/2}^q\tilde{y}_0^q}(\Omega)&\left|K^{\pi/2}_{yy}(\Omega)\right|^2S_{\tilde{y}_{\pi/2}^q\tilde{y}_{\pi/2}^q}(\Omega)
\end{array}
\right],
\end{split}
\end{equation}
where $S_{\tilde{y}_0^q\tilde{y}_0^q}(\Omega)=1, S_{\tilde{y}_0^q\tilde{y}_{\pi/2}^q}(\Omega)=S^*_{\tilde{y}_{\pi/2}^q\tilde{y}_0^q}(\Omega)=\alpha^2\chi_q(\Omega)$ and $S_{\tilde{y}_{\pi/2}^q\tilde{y}_{\pi/2}^q}(\Omega)=1+\alpha^4|\chi_q(\Omega)|^2+4\alpha^2|\chi_q(\Omega)|^2M\gamma_mk_BT$.
The determinant of this covariance matrix is
\be
{\rm det}\mathbf{V}(\Omega)=\left|K^{\pi/2}_{yy}(\Omega)\right|^2(1+4\alpha^2|\chi_q(\Omega)|^2M\gamma_mk_BT),
\ee
where ${\rm det}\mathbf{V}(\Omega)=1$ in standard quantum mechanics\,($K^{\pi/2}_{yy}(\Omega)=1$ since $R_q(\Omega)\rightarrow R_m(\Omega)$) with $T=0$. With sample parameters in Tab.\,\ref{tab:self_gravity}, we plot the ${\rm det}\mathbf{V}(\Omega)$ in Fig.\,\ref{fig:SN_detV},  which reveals that ${\rm det}\mathbf{V}(\Omega)$ can be less than one in the SN scenario at sufficiently low temperatures\,(in particular at $\Omega=\omega_q$), challenging the usual expectation of ${\rm det}\mathbf{V}(\Omega)\geq1$ for pure and mixed states.

In standard quantum mechanics, the Heisenberg uncertainty principle dictates that the covariance matrix of the outgoing field satisfies ${\rm det}[\mathbf{V}]\geq1$ or $S_{b_1b_1}S_{b_2b_2}-|S_{b_1b_2}|^2\geq1$\,\cite{Gibilisco2011}, reflecting that the optical field exchanges no net energy with the test mass. In the context of SN theory, however, when ${\rm det}\mathbf{V}(\Omega)<1$, this does not imply a violation of Heisenberg's principle. This is substantiated by the fact that both commutators, $[\hat b,\hat b^\dag]=1$ and $[\hat b^q,\hat b^{q\dag}]=1$, preserve the unitarity condition. Therefore, the covariance matrix under discussion does not precisely represent the quantum state of the light. The fundamental difference from quantum gravity is that $\hat b$ in SN theory is a quantum-classical mixed operator, incorporating a classical component from the SN term. Indeed, it is the covariance matrix of $\tilde{y}_\theta^q$, rather than $\tilde{y}_\theta$, that maintains the Heisenberg uncertainty principle by excluding the averaging over the SN term influenced by quantum trajectories.

Furthermore, the Wigner function of the outgoing field can be obtained from the covariance matrix in standard quantum optomechanics, which is called \emph{optical state tomography}\,\cite{Lvovsky2009,Miao2010}. For a Gaussian state, the cross-section of the Wigner function is an ellipse in the phase space of the optical state\,\cite{Kimble2001}, which can be determined by the optical quadrature spectrum measured at three different homodyne angles.  The foundation of this state tomography process is that, in standard quantum mechanics, different homodyne measurement angles correspond to the same state evolution. However, in the SN theory, measurements at different homodyne angles lead to different quantum trajectories, which source different evolution of the self-gravity potential and hence different mirror and optical state evolution.  Therefore, although Heisenberg picture analysis allows us to compute the covariance matrix in a similar way as standard quantum mechanics, the state tomography process in SN theory is subtle due to the intrinsic SN nonlinearity. Such a subtlety will also appear in the mutual gravity protocol as we shall see in the later sections.

\section{Mutual gravity protocol and entanglement}\label{sec5}
In this section, we explore the mutual gravity protocol to test the quantum nature of gravity\,\cite{Miao2020}. In particular, we further investigate the optical correlation in these experiments, and demonstrate that this correlation will constitute apparent entanglement that will be a false alarm of quantum gravity signature. In this setup, two optical fields, strengthened by cavities, track two test mass mirrors, which interact through mutual gravity. The equations describing the system's dynamics are given by:

\begin{equation}
\begin{split}
&\dot{\hat{x}}_{A/B}=\frac{\hat{p}_{A/B}}{M},\\
&\dot{\hat{p}}_{A/B}=-M\omega_q^2\hat{x}_A-\gamma_m\hat{p}_A+M\omega_g^2\langle\hat{x}_{B/A}\rangle_c+\alpha\hat{a}_{1{A/B}},\\
&\hat{b}_{\theta A/B}=\sin\theta\hat{a}_{2 A/B}+\cos\theta\hat{a}_{1 {A/B}}+\sin\theta\alpha\hat{x}_{A/B},
\end{split}
\end{equation}
where $\omega_g=\sqrt{2\mathcal{C}/M}$ and $\omega_q^2=\omega_m^2-\omega_g^2$.

Solving these motion equations results in:
\begin{equation}
\begin{split}
&\hat{b}_{\theta A/B}(t)=\cos\theta\hat{a}_{1 A/B}+\sin\theta\hat{a}_{2 A/B}\\
&+\sin\theta\alpha\int_{-\infty}^t\chi_q(t-t')(\hbar\hat{a}_{1 A/B}(t')+2\mathcal{C}\langle\hat{x}_{B/A}(t')\rangle_c)dt',
\end{split}
\end{equation}
where the quantum contribution \(\hat{b}_{\theta A/B}\) is defined by:
\begin{equation}
\begin{split}
\hat{b}^q_{\theta A/B}=&\cos\theta\hat{a}_{1 A/B}+\sin\theta\hat{a}_{2 A/B}\\
&+\sin\theta\alpha\int_{-\infty}^t\chi_q(t-t')\hbar\hat{a}_{1 A/B}(t')dt',
\end{split}
\end{equation}
and the projective measurement data of the outgoing light is:
\begin{equation}\label{eq:output_two_mirror}
\begin{split}
\tilde{y}_{\theta A/B}(t)=&\tilde{y}^q_{\theta A/B}(t)\\
&+2\mathcal{C}\sin\theta\alpha\int_{-\infty}^t\chi_q(t-t')\langle\hat{x}_{B/A}(t')\rangle_c dt',
\end{split}
\end{equation}
where $\tilde{y}^q_{\theta AB}$ pertains to the projective measurement data on $\hat{b}^q_{\theta AB}$. The primary challenge in the causal conditional prescription is extracting $\langle\hat{x}_{A/B}\rangle_c$ from $\tilde{y}_{\theta AB}$, which will be addressed using both the stochastic master equation filter method and the Wiener filter method.

\subsection{Filter functions}
\subsubsection{Wiener filter approach}
Following the procedure established for the self-gravity protocol, the displacement $\hat{x}(t)$ can be expressed as:
\begin{equation}
\hat{x}_{A/B}(t)=\int_{-\infty}^t\chi_q(t-t')(\hbar\alpha\hat{a}_{1A/B}(t')+M\omega_g^2\langle\hat{x}_{B/A}(t')\rangle_c)dt'.
\end{equation}
We separate $\hat{x}_{A/B}$ into $x^c_{A/B}$ and $\hat{x}^q_{A/B}$:
\begin{equation}
\begin{split}
\hat{x}^q_{A/B}(t)&=\int_{-\infty}^tdt'\chi_q(t-t')\hbar\alpha\hat{a}_{1A/B}(t'),\\
x^{\rm cl}_{A/B}(t)&=M\omega_g^2\int_{-\infty}^tdt'\chi_q(t-t')\langle\hat{x}_{B/A}(t')\rangle_c.
\end{split}
\end{equation}
Thus, $\langle\hat{x}^q_{A/B}\rangle\equiv\langle\hat{x}^q_{A/B}\rangle_c-x^{\rm cl}_{A/B}$ is related to $\tilde{y}^q_{\theta A/B}$ via the Wiener filter $K^{A/B}_{q}(t-t')$:
\begin{equation}
\langle\hat{x}^q_{A/B}(t)\rangle=\int_{-\infty}^tK^{A/B}_{q}(t-t')\tilde{y}^q_{\theta A/B}(t')dt'.
\end{equation}
In the SN theory's causal-conditional framework, the left and right optomechanical systems interact solely via the mutual gravitational influence of quantum trajectories. 

Therefore, $\tilde{y}^q_{A/B}$ is akin to the self-gravity protocol, aside from the modification of the resonance frequency by mutual gravity.
\begin{widetext}
\begin{figure*}
\centering
\includegraphics[width=0.9\textwidth]{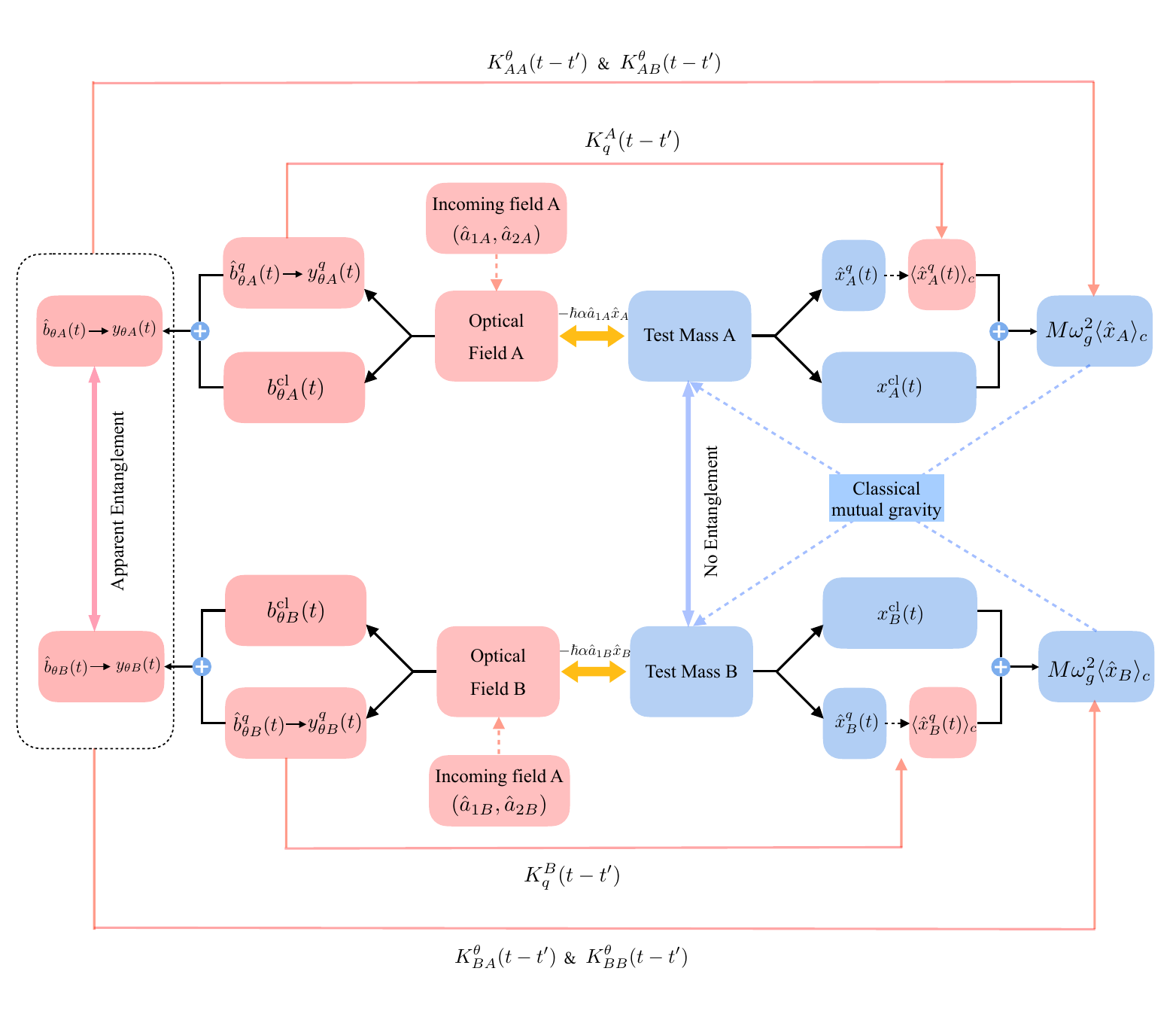}
\caption{Causal conditional dynamics as a Wiener filtering problem in the Heisenberg picture for the mutual-gravity protocol. In SN theory, the test masses interact through classical gravity, characterized by the c-number terms $M\omega_g^2\langle \hat x_{A/B}\rangle_c$. This implies that the quantum states of test masses A and B are not entangled. The appearance of apparent entanglement between the two outgoing fields arises from the interaction of the quantum trajectories of the test masses.}\label{fig:heisenberg_wiener_two}
\end{figure*}
\end{widetext}

For illustration, consider $\theta_A=\theta_B=\theta$. The frequency domain expression for $K^{A/B}_q(\Omega)$ is:
\begin{equation}\label{eq:filter_two}
K^{A/B}_q(\Omega)=\frac{\beta\beta_c+\Omega(\beta-\beta_c)+i\gamma_m\Omega-\omega_m^2}{\alpha\sin\theta(\beta-\Omega)(\Omega+\beta_c)}.
\end{equation} 
In the frequency domain, the conditional displacement of mirror A/B is:
\begin{equation}
\langle\hat{x}_{A/B}(\Omega)\rangle=K^{A/B}_q(\Omega)\tilde{y}^q_{\theta A/B}(\Omega)+M\omega_g^2\chi_q(\Omega)\langle\hat{x}_{A/B}(\Omega)\rangle,
\end{equation}
which leads to
\begin{equation}\label{eq:x}
\begin{split}
\langle\hat{x}_{A/B}(\Omega)\rangle &=\frac{R^2_q(\Omega)}{R_{+}(\Omega)R_{-}(\Omega)}K^{A/B}_q(\Omega)\tilde{y}^q_{\theta A/B}(\Omega)\\
&+\frac{\omega_g^2(\Omega)R_q(\Omega)}{R_{+}(\Omega)R_{-}(\Omega)}K^{B/A}_q(\Omega)\tilde{y}^q_{\theta B/A}(\Omega).
\end{split}
\end{equation}
Substituting into the output measurement data:
\begin{equation}\label{eq:y}
\tilde{y}_{\theta A/B}(\Omega)=\tilde{y}^q_{A/B}(\Omega)+\frac{M\omega_g^2\sin\theta\alpha}{R_q(\Omega)}\langle\hat{x}_{B/A}\rangle(\Omega),
\end{equation}
yields:
\begin{equation}\label{eq:y_AB_Wiener_theta}
\begin{split}
\tilde{y}_{\theta A/B}(\Omega)&=\left[1+\frac{\alpha\sin\theta\omega_g^4K^{A/B}_q(\Omega)}{R_{+}(\Omega)R_{-}(\Omega)}\right]\tilde{y}^q_{\theta A/B}(\Omega)\\
&+\left[\frac{\alpha\sin\theta\omega_g^2R_q(\Omega)K^{B/A}_q(\Omega)}{R_{+}(\Omega)R_{-}(\Omega)}\right]\tilde{y}^q_{\theta B/A}(\Omega).
\end{split}
\end{equation}
For the special case $\theta=\pi/2$:
\begin{equation}\label{eq:y_AB}
\begin{split}
\tilde{y}_{2 A/B}(\Omega)=&\left[1+\frac{\alpha\omega_g^4K^{A/B}_{q,\pi/2}(\Omega)}{R_{+}(\Omega)R_{-}(\Omega)}\right]\tilde{y}^q_{2 A/B}(\Omega)\\
&+\left[\frac{\alpha\omega_g^2R_q(\Omega)K^{B/A}_{q,\pi/2}(\Omega)}{R_{+}(\Omega)R_{-}(\Omega)}\right]\tilde{y}^q_{2 B/A}(\Omega).
\end{split}
\end{equation}

By employing Eq.\,\eqref{eq:filter_two}, we find:
\begin{equation}\label{eq:y_2AB_mutual_Wiener}
\begin{split}
&\tilde{y}_{2 A/B}(\Omega)=\\
&\left[1+\frac{\omega_g^4}{R_{+}(\Omega)R_{-}(\Omega)}\frac{(\beta-\Omega)(\beta_c+\Omega)-R_m(\Omega)}{(\beta-\Omega)(\beta_c+\Omega)}\right]\tilde{y}^q_{2 A/B}(\Omega)\\
&+\left[\frac{\omega_g^2R_q(\Omega)}{R_{+}(\Omega)R_{-}(\Omega)}\frac{(\beta-\Omega)(\beta_c+\Omega)-R_m(\Omega)}{(\beta-\Omega)(\beta_c+\Omega)}\right]\tilde{y}^q_{2 B/A}(\Omega).
\end{split}
\end{equation}
Calculations in Appendix A.2 confirm that $\tilde{y}_{2 A/B}(\Omega)$\,(Eq.\eqref{eq:y_2AB_mutual_Wiener}) matches precisely with the result obtained by the stochastic master equation method\,(Eq.\eqref{eq:y_2AB_mutual_sme}).

\subsubsection{Stochastic Master equation approach}
Firstly, we calculate the filter function based on the long-term behavior of the stochastic master equation. To extract $\langle\hat{x}_{A/B}\rangle_c$ from $\tilde{y}_{\theta A/B}$, assuming $\theta=\pi/2$, we propose a two-channel filter matrix:

\begin{equation}
\left[
\begin{array}{c}
\langle\hat{x}_A(t)\rangle_c\\
\langle\hat{x}_B(t)\rangle_c
\end{array}
\right]
=\int_{-\infty}^tdt'
\left[
\begin{array}{cc}
K_{AA}(t-t')&K_{AB}(t-t')\\
K_{BA}(t-t')&K_{BB}(t-t')\\
\end{array}
\right]
\left[
\begin{array}{c}
\tilde{y}_{2A}(t')\\
\tilde{y}_{2B}(t')
\end{array}
\right].
\end{equation}

Solving these filter functions involves the evolution equations of the conditional expectation value $\langle\hat{x}_{A/B}\rangle_c$:

\begin{equation}\label{eq:expectation_two_mirrors}
d
\left[
\begin{array}{c}
\langle\hat{x}_A\rangle_c\\
\langle\hat{p}_A\rangle_c\\
\langle\hat{x}_B\rangle_c\\
\langle\hat{p}_B\rangle_c
\end{array}
\right]
=
\mathbb{M}
\left[
\begin{array}{c}
\langle\hat{x}_A\rangle\\
\langle\hat{p}_A\rangle\\
\langle\hat{x}_B\rangle\\
\langle\hat{p}_B\rangle
\end{array}
\right]
+
2\alpha \left[
\begin{array}{c}
V^A_{xx}\tilde{y}_{2A}\\
V^A_{xp}\tilde{y}_{2A}\\
V^B_{xx}\tilde{y}_{2B}\\
V^B_{xp}\tilde{y}_{2B}
\end{array}
\right],
\end{equation}
where 
\be
\mathbb{M}=
\left[
\begin{array}{cccc}
-2\alpha^2V^A_{xx}&1/M&0&0\\
-M\omega_{q}^2-2\alpha^2V^A_{xp}&-\gamma_m&M\omega_g^2&0\\
0&0&-2\alpha^2V^B_{xx}&1/M\\
M\omega_g^2&0&-M\omega_{q}^2-2\alpha^2V^B_{xp}&-\gamma_m
\end{array}
\right]
\ee
and we have reexpress all $dW_{A/B}$ in terms of $\tilde{y}_{\theta A/B}$.

In a symmetric configuration, where both optomechanical cavities have matching parameters, we have $V^A_{xx}=V^B_{xx}=V_{xx}$ and $V^A_{xp}=V^B_{xp}=V_{xp}$. By solving Eq.\,\eqref{eq:expectation_two_mirrors}, the two-channel filter function in the frequency domain is:
\begin{equation}
\begin{split}
K_{AA/BB}(\Omega)&=\frac{A(\Omega)}{\alpha(2A(\Omega)+R^{-}(\Omega))}+\frac{A(\Omega)}{\alpha(2A(\Omega)+R^{+}(\Omega))},\\
K_{AB/BA}(\Omega)&=\frac{A(\Omega)}{\alpha(2A(\Omega)+R^{-}(\Omega))}-\frac{A(\Omega)}{\alpha(2A(\Omega)+R^{+}(\Omega))},
\end{split}
\end{equation}
where
\begin{equation}
\begin{split}
&A(\Omega)=V_{xx}\alpha^2(V_{xx}\alpha^2+\gamma_m-i\Omega),\\
&R^{-}(\Omega)=\omega_q^2-\omega_g^2-\Omega^2-i\gamma_m\Omega,\\
&R^{+}(\Omega)=\omega_q^2+\omega_g^2-\Omega^2-i\gamma_m\Omega.
\end{split}
\end{equation}

The frequency domain measurement record, as stated in Eq.\,\eqref{eq:output_two_mirror}, is now:
\begin{equation}
\begin{split}
&\left[
\begin{array}{c}
\tilde{y}_{2A}(\Omega)\\
\tilde{y}_{2B}(\Omega)
\end{array}
\right]=
\\
&\left[\mathbf{1}-M\omega_g^2\alpha\chi_q(\Omega)
\left(
\begin{array}{cc}
K_{BA}(\Omega)&K_{BB}(\Omega)\\
K_{AA}(\Omega)&K_{AB}(\Omega)
\end{array}
\right)\right]^{-1}
\left[
\begin{array}{c}
\tilde{y}^q_{2A}(\Omega)\\
\tilde{y}^q_{2B}(\Omega)
\end{array}
\right].
\end{split}
\end{equation}

This simplifies to:
\begin{equation}\label{eq:y_2AB_mutual_sme}
\begin{split}
\tilde{y}_{2A/B}=&\left[1+\frac{2\omega_g^2}{R^{+}(\Omega)R^{-}(\Omega)}\frac{\omega_g^2A(\Omega)}{2A(\Omega)+R^q(\Omega)}\right]\tilde{y}^q_{2A/B}\\
&+\left[\frac{2\omega_g^2}{R^{+}(\Omega)R^{-}(\Omega)}\frac{A(\Omega)R^q(\Omega)}{2A(\Omega)+R^q(\Omega)}\right]\tilde{y}^q_{2B/A}.
\end{split}
\end{equation}

\subsection{apparent entanglement of outgoing light fields}
As mentioned in the Introduction, probing gravity-induced entanglement (GIE) can offer strong evidence for quantum gravity\,\cite{Bose2017,krisnanda2020,Miao2020,Marshman2020}. In our mutual-gravity optomechanical protocol, such entanglement could manifest in the outgoing light, as suggested by\,\cite{Miao2020}. Paper I indicated that correlations in the outgoing light can also occur in the SN theory under the casual conditional framework, where mutual gravity is classical, and these correlations are nearly indistinguishable from those in quantum gravity. It remains unclear whether the complete entanglement structure is similarly indistinguishable.
  
\begin{table}[h!]
    \centering
    \begin{tabular}{|c|c|c|}
    \hline
Parameters&Symbol&Value\\
\hline
Mirror mass&$M$&$10^{-3}$\,kg\\
\hline
Mirror bare frequency&$\omega_m/(2\pi)$&$0.5\,{\rm Hz}$\\
\hline
SN frequency&$\omega_{g}/(2\pi)$&$2\times10^{-4}\,{\rm Hz}$\\
\hline
Quality factor&$Q_m$&$3\times10^7$\\
\hline
Mechanical damping&$\gamma_m/(2\pi)$&$1.67\times10^{-8}\,{\rm Hz}$\\
\hline
Optical wavelength&$\lambda$&$1064\,{\rm nm}$\\
\hline
Cavity Finesse&$\mathcal{F}$&$4000$\\
\hline
Intra-cavity power&$P_{\rm cav}$&$2000\,{\rm W}$\\
\hline
    \end{tabular}
   \caption{Sample parameters for the mutual-gravity optomechanical protocol.}\label{tab:mutual_gravity_parameter}
\end{table}

Operationally, imagine that future experimentalists conduct such an optomechanical GIE experiment, obtain data from the projective measurement of the phases of the two outgoing fields, and subsequently process this data to compute an entanglement indicator such as the logarithmic negativity $\epsilon_N$ as defined in Eq.\eqref{eq:logarithmic_negativity}. Can a non-negative value of $\epsilon_N$, extracted in this manner, serve as definitive evidence of quantum gravity? To address this question, it is necessary to perform a similar calculation for the mutual gravity protocol within the SN theory framework.

\begin{figure*}
\centering
\includegraphics[width=0.8\textwidth]{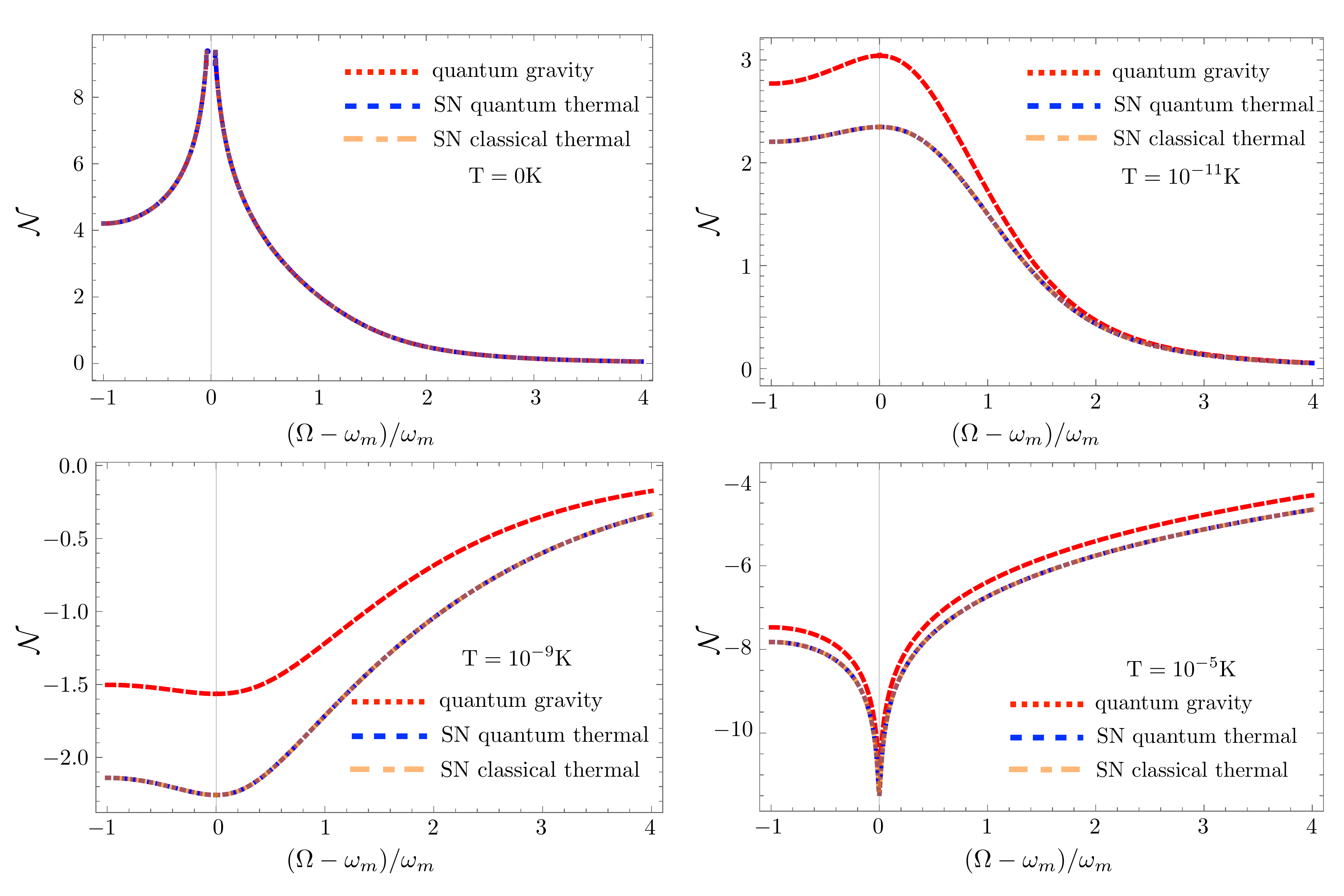}
\caption{Apparent entanglement of the outgoing optical fields in the mutual gravity protocol, where $\mathcal{N}$ is defined from the logarithmic negativity $\epsilon_N={\rm max}\{\mathcal{N},0\}$ and $\mathcal{N}=-(1/2){\rm ln}\left[\left(\Sigma-\sqrt{\Sigma^2-4{\rm det}\sigma}\right)/2\right]$ where $\mathcal{N}>0$ means the exist of entanglement. The blue-dashed curve is the $\mathcal{N}$ of the outgoing optical state in the SN theory at different frequencies, where the thermal noise is treated quantum mechanically. For comparison, the orange and red curves are the $\mathcal{N}$ in the SN theory when the thermal noise is treated classically and in the quantum gravity theory, respectively. The details of the influence of classical thermal noise are discussed in Appendix.\,\ref{sec:classical_thermal} }\label{fig:entanglement_two_mirrors}
\end{figure*}

The Heisenberg-picture method offers a means to perform this calculation in the SN theory. Specifically, we need to compute the full covariance matrix of the outgoing light\,\eqref{eq:covariance_matrix_general}, focusing on the non-diagonal elements of matrices $\sigma_A=\sigma_B$ and $\sigma_{AB}$. Using Eq.\eqref{eq:y_AB_Wiener_theta}, we have:
\begin{equation}\label{eq:y_AB_Wiener_12}
\begin{split}
\tilde{y}_{1 A/B}(\Omega)=&\tilde{y}^q_{1 A/B}(\Omega),\\
\tilde{y}_{2A/B}(\Omega)=&\mathcal{T}_{AA/BB}(\Omega)\tilde{y}^q_{2 A/B}(\Omega)+\mathcal{T}_{AB/BA}(\Omega)\tilde{y}^q_{2 B/A}(\Omega),
\end{split}
\end{equation}
where
\be
\begin{split}
&\mathcal{T}_{AA}(\Omega)=\mathcal{T}_{BB}(\Omega)=1+\frac{\alpha\omega_g^4K^{\pi/2 \rm Wiener}_{A/B}(\Omega)}{R_{+}(\Omega)R_{-}(\Omega)},\\
&\mathcal{T}_{AB}(\Omega)=\mathcal{T}_{BA}(\Omega)=\frac{\alpha\omega_g^2R_q(\Omega)K^{\pi/2 \rm Wiener}_{B/A}(\Omega)}{R_{+}(\Omega)R_{-}(\Omega)}.
\end{split}
\ee
The covariance matrices $\sigma_A$ and $\sigma_{AB}$ are then:
\begin{equation}
\begin{split}
\sigma_{A/B}=\left[
\begin{array}{cc}
1&\mathcal{T}^*_{AA}\alpha^2\chi^*_q\\
\mathcal{T}_{AA}\alpha^2\chi_q&(|\mathcal{T}_{AA}|^2+|\mathcal{T}_{AB}|^2)\left(1+\alpha^4|\chi_q|^2\right)
\end{array}
\right],
\end{split}
\end{equation}
and
\begin{equation}
\sigma_{AB}=\sigma_{BA}
\left[
\begin{array}{cc}
0&\mathcal{T}^{*}_{AB}\alpha^2\chi^*_q\\
\mathcal{T}_{AB}\alpha^2\chi_q&2{\rm Re}[\mathcal{T}^{*}_{AA}\mathcal{T}_{BA}]\left(1+\alpha^4|\chi_q|^2\right)
\end{array}
\right].
\end{equation}
Utilizing the sample parameters outlined in Table \ref{tab:mutual_gravity_parameter}, we plotted the logarithmic negativity as shown in Figure \ref{fig:entanglement_two_mirrors}, comparing these findings to those derived from the quantum gravity framework. The results indicate that at zero temperature, both the SN theory and quantum gravity theory predict identical values for negativity. However, when accounting for the influence of a quantum thermal bath, this equivalence is quantitatively disrupted. Notably, the weak, apparent entanglement ($\epsilon_N>0$) appears only under very low temperatures, as even a minor thermal effect can eliminate this apparent entanglement.

The influence of a quantum thermal bath on apparent entanglement is nuanced. In the SN theory, the expectation value of the displacement of mirrors A/B under the influence of a quantum bath remains zero, and thus, does not directly contribute to the classical gravity that impacts the motion of mirrors B/A. In contrast, within the framework of quantum gravity, the thermal bath can directly influence $\hat x_{A/B}$, subsequently affecting the evolution of the operator $\hat x_{B/A}$. Nevertheless, the thermal noise indirectly influences the mutual gravity protocol in SN theory by infiltrating the quantum trajectory of mirrors A/B through its additional contribution to the conditional variance $V^{A/B}_{xx}$. This differing impact of thermal noise on the dynamics of mutual gravity protocols in quantum gravity and SN theories accounts for the variations observed in negativity.

Finally, our analysis reveals that classical gravity can induce a apparent entanglement in the outgoing optical field within the SN theory. To correctly grasp the implications of this result for the GIE experiment, in particular for SN theory, it is beneficial to revisit the framework of the GIE experiment.

The conceptual foundation of the GIE experiment is based on the premise that quantum Newtonian mutual gravity can induce entanglement between two massive mirrors. However, experimentally accessing information about the mirror states can only be achieved through a detector interacting with the mirrors, which, in the context of the optomechanical protocol, is the optical field. Consequently, the entanglement between the mirrors is transferred to the optical field via optomechanical interactions, allowing it to manifest as entanglement in the optical domain.

Suppose some future experimentalists and theorists collaborated and designed such an experiment with the parameters listed in Tab.\,\ref{tab:mutual_gravity_parameter}. Then the experimentalist built and run the experimental apparatus, and collected the optical quadrature data of outgoing light A/B:  $\tilde y_{\theta A/B}(t)$,
\be\label{eq:y_theta_A/B_QG}
\tilde y_{\theta A/B}(t)=\sin\theta \tilde y_{1 A/B}(t)+\cos\theta \tilde y_{2 A/B}(t).
\ee 
These data allow the experimentalist to \emph{directly compute} their correlation function $C_{\tilde y_{\theta A/B}\tilde y_{\theta A/B}}(\tau)$, satisfying
\be
\begin{split}
C_{\tilde y_{\theta A/B}\tilde y_{\theta A/B}}(\tau)=&\sin^2\theta C_{\tilde y_{1 A/B}\tilde y_{1 A/B}}(\tau)+\cos^2\theta C_{\tilde y_{2 A/B}\tilde y_{2 A/B}}(\tau)\\
&+\sin \theta\cos \theta C_{\tilde y_{1 A/B}\tilde y_{2 A/B}}(\tau).
\end{split}
\ee
Measuring three different quadratures allows us to calculate $C_{\tilde y_{1 A/B}\tilde y_{1 A/B}}(\tau),C_{\tilde y_{2A/B}\tilde y_{2 A/B}}(\tau)$ and $C_{\tilde y_{1 A/B}\tilde y_{2 A/B}}(\tau)$, which the can be Fourier transformed to obtain the power spectrum\,(note that $S_{\tilde y_{1 A/B}\tilde y_{1 A/B}}(\Omega)$ and $S_{\tilde y_{2 A/B}\tilde y_{2 A/B}}(\Omega)$ are real, and the cross-spectrum can be complex) and thereby the covariance matrix.

However, the above prescription is valid only for quantum gravity, where everything is linear. The nonlinearity of the SN theory adds additional subtleties to this prescription. Note that the input-output relation in SN theory is given by Eq.\,\eqref{eq:y_AB_Wiener_theta} (or \eqref{eq:y_AB_Wiener_12}), which can be formally written in the time domain as:
\be\label{eq:y_theta_A/B_SN}
\begin{split}
\tilde{y}_{\theta A/B}(t)=\int^t_{-\infty}dt'\mathcal{T}^\theta_{AA/BB}(t-t')[\sin\theta,\cos\theta]
\left[
\begin{array}{c}
\tilde{y}^q_{1A/B}(t')\\
\tilde{y}^q_{2A/B}(t')
\end{array}
\right]\\
+\int^t_{-\infty}dt'\mathcal{T}^\theta_{AB/BA}(t-t')
[\sin\theta,\cos\theta]
\left[
\begin{array}{c}
\tilde{y}^q_{1B/A}(t')\\
\tilde{y}^q_{2B/A}(t')
\end{array}
\right]
\end{split}
\ee
Since the $\mathcal{T}^\theta_{AA/BB}(\Omega),\mathcal{T}^\theta_{AB/BA}(\Omega)$ depend on $\theta$, these relations become highly nonlinear. Therefore, we are not able to solve the cross-spectrum and the corresponding covariance matrix using the previous prescription. Moreover, if Nature followed SN theory while the experimentalists did not know and believed in QG, then the experimentalists analyzed their data using Eq.\,\eqref{eq:y_theta_A/B_QG} rather than Eq.\,\eqref{eq:y_theta_A/B_SN}. Then, this prescription will lead to a $\theta$-dependent negativity, which physically makes no sense. Finally, although the apparent entanglement can be calculated for SN theory, it remains an open question of finding a proper experimental state tomography method for the outgoing optical state.
 
\section{Discussion and Conclusion}\label{sec5}

In conclusion, this study introduces a Heisenberg-picture approach for analyzing the Schrödinger-Newton theory within the causal conditional framework, which incorporates the impact of continuous quantum measurement on system dynamics. We have demonstrated the equivalence between this Heisenberg-picture method and the Schrödinger-picture method previously employed in Paper I. Utilizing these theoretical tools, we further examined two optomechanical protocols aimed at testing the SN theory and probing gravity-induced entanglement (GIE) by analyzing the correlation matrix of the output optical fields. Our findings indicate that classical gravity can lead to apparent entanglement between the two output light fields, potentially serving as a false alarm in experiments designed to demonstrate the quantum nature of Newtonian gravity through optical field entanglement.

Significantly, the SN dynamics in the mutual gravity protocol can be interpreted as a \emph{Local Operations and Classical Communications (LOCC)}\,\cite{chitambar2014,Fan2004} interaction process. Building on the findings from Paper I, this study provides a concrete example of how LOCC-type interactions can imitate quantum interactions, potentially leading to false alarm of quantum effects. Intriguingly, recent work\,\cite{Lami2024} by the Plenio group addresses this issue by developing a methodology to test the quantumness of gravity without relying on entanglement. In their research, they quantify the maximum fidelity with which any classical interaction, such as a LOCC-type interaction, can replicate a given quantum interaction. It is important to note that the fidelity between the conditional mirror states prepared under continuous measurement in both quantum gravity and SN theory is expected to adhere to Plenio's upper bound, a topic that will be explored in future work.

Furthermore, several points need to be discussed to provide some insights into understanding this apparent entanglement.

\emph{Unentangled mirrors---}In quantum gravity, the entanglement of the two optical fields can be attributed back to the entanglement of the two mirror oscillators, induced by the quantum Newtonian mutual gravity\,\cite{Miao2020}. In contrast, there is \emph{no entanglement} between the mirror oscillators since the classical Newtonian gravity can not transfer quantum information\,\cite{Liu2023}. The ``apparent entanglement" of the two outgoing light fields exists only because the SN dynamics of mirror B depend on the measurement outcome of mirror A in the causal conditional formalism, which is a feature of the SN theory as a nonlinear quantum mechanics. However, as we have shown in Fig.\,\ref{fig:heisenberg_wiener_two}, this apparent entanglement is sufficient to be a false alarm for quantum gravity.

\emph{Quantum state---}Heisenberg picture analysis of the self-gravity and mutual gravity optomechanical protocol reveals interesting features of the light field interacting with the test mass state in the Schr{\"o}dinger-Newton theory. In particular, for the mutual-gravity case, we obtained a apparent entanglement in the SN theory which is almost indistinguishable from the real entanglement if the gravity is quantum.  Entanglement measures the non-separable feature of the quantum state, therefore the quantum state of the optical fields here deserves a further discussion.

In standard quantum mechanics\,(or when the gravity is quantum), the quantum state\,(or the wavefunction) is considered to be the physical reality according to the standard interpretation. In a unitary process, the quantum state evolves linearly; while in a quantum measurement process, the quantum state manifests its statistical characters.  However, experimentally extracting the quantum information in the quantum state generally requires a quantum measurement process that collapses the quantum state. The experimentalists collect the data from the measurement ensemble, reconstruct the data statistics, and finally reconstruct the quantum state. This process is called \emph{quantum tomography}. In standard quantum mechanics, suppose we have a state $|\psi(t)\rangle$ evolved unitarily from an initial state $|\psi(0)\rangle$. The state obtained by quantum tomography on the $|\psi(t)\rangle$ is the same as $|\psi(t)\rangle$ in standard quantum mechanics.

In the SN theory, subtleties arise due to its inherent nonlinearity. For an initially unentangled joint optical state given by $|\psi_{\rm opt}(0)\rangle=|0\rangle_A\otimes|0\rangle_B$, the evolution of the state will depend on the measurement outcomes $\tilde y_{\theta A},\tilde y_{\theta B}$. This dependence occurs because these measurement outcomes determine the conditional mechanical state, which in turn modulates the incoming optical field. Consequently, the evolution operator for the optical state $|\psi_{\rm opt}(t)\rangle$ incorporates $\tilde y_{\theta A/B}$. Intriguingly, if the standard approach is employed to reconstruct the optical quantum state from the covariance matrix—akin to quantum state tomography—the entanglement attribute will manifest as positive negativity, provided thermal effects are sufficiently minimized. Conversely, a direct derivation of the causal conditional dynamics from first principles reveals that such perceived entanglement is only apparent, and no actual entanglement exists in the state $|\psi_{\rm opt}(t)\rangle$. This discrepancy highlights a break in the equivalency that holds in standard quantum mechanics, signaling a subtle complexity in interpreting the wave function within the SN framework.

Regarding the experiment designed to probe quantum gravity such as GIE-experiment,  the potential false alarm found in this work shows that carefully designing the experiment for circumventing the effect of quantum trajectory induced by continuous quantum measurement is important, which we will leave for future work.

\acknowledgments
Y.M. thanks Professor Chunnong Zhao for the useful discussions. Y.M. is supported by the National Key R$\&$D Program of China ``Gravitational Wave Detection" (Grant No. 2023YFC2205801), National Natural Science Foundation of China under Grant No.12474481, and the start-up funding provided by Huazhong University of Science and Technology. Y. C. is supported by the Simons Foundation\,(Award Number 569762).

\appendix
\section{Equivalence of Wiener filter method and the stochastic master equation method in standard quantum mechanics}
In this section, we discuss two methods that can obtain the conditional expectation value in standard quantum mechanics, using an exemplary optomechanical system. Firstly, we present the Wiener method and its application in the optomechanism system. Secondly, we present the stochastic master equation method and calculate the Kalman filter to obtain the steady-state conditional expectation value. Lastly, we prove that these methods are exactly equivalent in the steady state of a continuous measurement process.

\subsection{The Wiener filter method}\label{subsec:Wiener_filter}
A simple optomechanical system has the following equations of motion:
\begin{equation}
\begin{split}
M\ddot{\hat{x}}&=-M\omega_m^2\hat{x}-\gamma M\dot{\hat{x}}+\alpha\hat{a}_1\\
\hat{b}_\theta&=\cos\theta\hat{a}_1+\sin\theta\hat{a}_2+\sin\theta\alpha\hat{x}
\end{split}
\end{equation}
The minimum means square estimator\,(MMSE) on the measurement record\,(on the $\hat b_\theta$) database $\tilde{y}_\theta$ gives the conditional expectation of mechanical displacement $\langle x(t)\rangle_c$ as follows.
The estimator of the mechanical displacement can be constructed with the filter function $K(t-t')$,
\begin{equation}
x_{\rm est}(t)=\int_{-\infty}^tK(t-t')\tilde y_\theta(t')dt',
\end{equation}
where the MMSE requests that the filter function $K(t-t')$ satisfies the variational principle:
\be
\frac{\delta}{\delta K}\left\langle\left(\hat x(t)-\int_{-\infty}^tK(t-t')\tilde y_\theta(t')dt'\right)^2\right\rangle=0.
\ee
This variational principle leads to an integral equation that must be satisfied by the filter called the Wiener filter $K_w(t-t')$:
\be
C_{x\tilde y_\theta}(t-t'')=\int^t_{-\infty}dt'K_w(t-t')C_{\tilde y_\theta\tilde y_\theta}(t'-t''),\quad t''\leq t,
\ee
where $C_{AB}$ is the correlation function of quantities $A,B$. This result can also be understood as the error of the estimator $\hat x(t)-x_{\rm est}(t)$ have no correlation with the past measurement data $\tilde y_\theta(t'<t)$.
The Wiener filter can be computed by the Wiener-Hopf equation in the frequency domain. 

This integral equation can be solved via the Wiener-Hopf method:
\be\label{eq:Wiener-Hopf}
K_{\rm Wiener}(\Omega)=\frac{1}{\phi_+(\Omega)}\left[\frac{S_{x\tilde y_\theta}(\Omega)}{\phi_-(\Omega)}\right]_+
\ee
In the above formula, we decompose the outgoing spectrum as $S_{\tilde y_\theta\tilde y_\theta}(\Omega)=\phi_+(\Omega)\phi_-(\Omega)$, where the $\phi_{\pm}(\Omega)$ and their inverse are analytical in the upper/lower-half complex $\Omega$-plane. The $[...]_+$ also means that we only take the part which is analytical in the upper complex $\Omega$-plane. In order to obtain the Wiener-filtering function, we need to have $\phi_\pm(\Omega)$ and $S_{x\tilde y_\theta}(\Omega)$.

First, it is straightforward to show that
\begin{equation}
\begin{split}
S_{\tilde y_\theta \tilde y_\theta}&=\frac{\sin^2\theta\alpha^4+M\alpha^2\sin{2\theta}(\omega_m^2-\Omega^2)}{M^2|\omega_m^2-\Omega^2-i\gamma_m\Omega|^2}+1,
\end{split}
\end{equation}
which can be rearranged as
\begin{equation}
S_{\tilde y_\theta \tilde y_\theta}=\frac{(\Omega-\beta)(\Omega+\beta)(\Omega-\beta_c)(\Omega+\beta_c)}{(\Omega-\eta)(\Omega+\eta)(\Omega-\eta_c)(\Omega+\eta_c)},
\end{equation}
where 
\be
\begin{split}
&\beta=a+ib,\quad\beta_c=a-ib,\\
&\eta=\frac{\sqrt{-\gamma_m^2+4\omega_m^2}}{2}-i\frac{\gamma_m}{2},
\end{split}
\ee
with
\begin{equation}
\begin{split}
a=\frac{\sqrt{a_1+\sqrt{a_2}}}{2},\quad b=-\frac{\sqrt{-a_1^2-a_2}}{4a},
\end{split}
\end{equation}
and
\begin{equation}
\begin{split}
&a_1=-\gamma_m^2+2\omega_m^2+\Lambda^2\sin2\theta,\\
&a_2=2\Lambda^4(1-\cos2\theta)+4\omega_m^2(\omega_m^2+\Lambda^2\sin2\theta).
\end{split}
\end{equation}
Then according to the definition of $\phi_\pm(\Omega)$, we have:
\begin{equation}\label{eq:phi_standard_qm}
\begin{split}
\phi_{+}(\Omega)&=\frac{(\Omega-\beta)(\Omega+\beta_c)}{(\Omega-\eta)(\Omega+\eta_c)},\\
\phi_{-}(\Omega)&=\frac{(\Omega-\beta_c)(\Omega+\beta)}{(\Omega-\eta_c)(\Omega+\eta)}.
\end{split}
\end{equation}

Secondly, we have
\begin{equation}
\begin{split}
S_{x\tilde y_\theta}&=\cos\theta\alpha S_{xy_1}+\sin\theta\alpha S_{xx}\\
&=\frac{\sin\theta\alpha^3-\cos\theta\alpha M(\Omega+\eta)(\Omega-\eta_c)}{M^2(\Omega-\eta)(\Omega+\eta)(\Omega-\eta_c)(\Omega+\eta_c)}.
\end{split}
\end{equation}
This $S_{x\tilde y_\theta}$, together with Eq.\eqref{eq:phi_standard_qm} and some tedious alegbras, leads to,
\begin{equation}
\left[\frac{S_{xb_2}}{\phi_{-}}\right]_{+}=\frac{\beta\beta_c+\Omega(\beta-\beta_c)+i\gamma_m\Omega-\omega_m^2}{\alpha\sin\theta((\eta-\Omega)(\Omega+\eta_c))}.
\end{equation}
Therefore, by the definition Eq.\,\eqref{eq:Wiener-Hopf}, it is straightforward to obtain the Wiener filter function:
\begin{equation}\label{eq:Wiener_filter_standard_qm}
K_{\rm Wiener}(\Omega)=\frac{\beta\beta_c+\Omega(\beta-\beta_c)+i\gamma_m\Omega-\omega_m^2}{\alpha\sin\theta(\beta-\Omega)(\Omega+\beta_c)}.
\end{equation}

\subsection{The filter derived from the stochastic master equation and the equivalency}\label{subsec:filter_w}
The method and the formal solution is exactly the same as in the main text. The obtained Kalman filter function at the long-time tail can be written as:
\begin{equation}
K_{\rm Kalman}(\Omega)=\frac{A_m(\Omega)}{\alpha\sin\theta(A_m(\Omega)-R_m(\Omega))}.
\end{equation}
Where,
\begin{equation}\label{eq:A_mR_m}
\begin{split}
A_m(\Omega)=&V_{xx}\alpha^2(V_{xx}\alpha^2\cos{2\theta}-V_{xx}\alpha^2-2\gamma_m+2i\Omega)\sin^2\theta\\&-\alpha^2\sin\theta\cos\theta/M,\\
R_m(\Omega)&=\omega_m^2-\Omega^2-i\gamma_m\Omega.
\end{split}
\end{equation}
The only difference is the second order moment $V_{xx}$ here takes the following form:
\begin{equation}
\begin{split}
&V_{xx}=\frac{-\gamma_m}{2\alpha^2\sin^2\theta}+\\
&\frac{\sqrt{\gamma_m^2-2\omega_m^2-\Lambda^2\sin2\theta+2\sqrt{\omega_m^4+\Lambda^4\sin^2\theta+\Lambda^2\omega_m^2\sin2\theta}}}{2\alpha^2\sin^2\theta},
\end{split}
\end{equation}
which is obtained from the Riccati equation for an optomechanical continuous measurement in the standard quantum mechanics. This $V_{xx}$ has a more compact form
if we use the definition of $\beta,\beta_c$:
\be
V_{xx}=\frac{-\gamma_m}{2\alpha^2\sin^2\theta}+\frac{i(\beta-\beta_c)}{2\alpha^2\sin^2\theta}.
\ee
Then the filter based on the master equation can be rearranged as:
 \begin{equation}\label{eq:master_filter}
 \begin{split}
 &K_{\rm Kalman}(\Omega)=\frac{1}{\alpha\sin\theta}\times\\
 &\frac{(\beta-\beta_c+i\gamma_m)(\beta-\beta_c-i\gamma_m-2\Omega)-\alpha^2\sin2\theta/M}{(\beta-\Omega)^2+(\beta_c+\Omega)^2-2\beta\beta_c+\gamma_m^2-2\omega_m^2-\alpha^2\sin2\theta/M}.
 \end{split}
 \end{equation}
 Using the definition of $\beta$ and $\beta_c$, we can derive the following useful relation:
 \begin{equation}\label{eq:wm_lambda}
 \begin{split}
 \omega_m&=\sqrt{\frac{\beta^2+\beta_c^2+\gamma_m^2-\Lambda^2\sin2\theta}{2}},\\
 \Lambda&=\sqrt{\frac{\beta^2+\beta_c^2+\gamma_m^2-2\omega_m^2}{\sin2\theta}}.
 \end{split}
 \end{equation}
 Combining Eq.\,\ref{eq:master_filter} and Eq.\,\ref{eq:wm_lambda} leads to,
 \begin{equation}
 K_{\rm Kalman}(\Omega)=\frac{\beta\beta_c+\Omega(\beta-\beta_c)+i\gamma_m\Omega-\omega_m^2}{\alpha\sin\theta(\beta-\Omega)(\Omega+\beta_c)},
 \end{equation}
 which is exactly equal to Eq.\eqref{eq:Wiener_filter_standard_qm}.
Therefore, for optomechanical continuous measurement in standard quantum mechanics, the Wiener filter method and the master equation method are exactly equivalent.

\section{Prove the equality satisfied by the spectrum \texorpdfstring{$S_{\tilde y^q_\theta\tilde y^q_\theta}$ in SN theory}{}.}
Here we present the proof of an equality that is used in the main text to simplify the formula.

Direct calculation of the $S_{\tilde y^q_\theta\tilde y^q_\theta}$ from the input-output relation $b_\theta^q$ leads to:
\begin{equation}\label{eq:Appendix_S_yq_self}
\begin{split}
S_{\tilde{y}^q_{\theta}\tilde{y}^q_{\theta}}=\frac{M^2|R_q(\Omega)|^2+\alpha^4\sin^2\theta+M\alpha^2\sin2\theta(\omega_q^2-\Omega^2)}{M^2|R_q(\Omega)|^2}.
\end{split}
\end{equation}
Using Eq.A16-18, and replace the $\omega_m$ in Eq.A17 by $\omega_q$, we obtain,
\begin{equation}\label{eq:A_Omega}
A^\theta_q(\Omega)=\frac{M(\beta-\beta_c+i\gamma_m)(\beta-\beta_c-i\gamma_m-2\Omega)-\alpha^2\sin(2\theta)}{2M}.
\end{equation}
On the other hand, using Eq.\,\ref{eq:A_Omega} and $R_q(\Omega)=\omega_q^2-\Omega^2-i\gamma_m\Omega$, we have,

\begin{equation}\label{eq:Apendix_S_y0}
\begin{split}
|A^\theta_q(&\Omega)-R_q(\Omega)|^2=\\
&|R_q(\Omega)|^2-\frac{(\beta-\beta_c)^2(\beta+\beta_c)^2}{4}\\
&+\frac{\gamma_m^2(\gamma_m^2-4\omega_q^2)}{4}-\frac{\alpha^2\gamma_m^2\sin(2\theta)}{2M}\\
&+\frac{\alpha^2(\omega_q^2-\Omega^2)\sin(2\theta)}{M}+\frac{\alpha^4\sin^2(2\theta)}{4M^2}.
\end{split}
\end{equation}
During the above derivation, the following relation has been used:
\be
\beta^2+\beta_c^2=-\gamma_m^2+2\omega_q^2+\frac{\alpha^2}{M}\sin(2\theta).
\ee 
Using the definition of $\beta$ and $\beta_c$, we have:
\begin{equation}\label{eq:Apendix_S_y0_1}
\begin{split}
(\beta-\beta_c)^2(\beta+\beta_c)^2=&\gamma_m^4-4\gamma_m^2\omega_q^2-2\frac{\alpha^2\gamma_m^2\sin(2\theta)}{M}\\
&-4\frac{\alpha^4\sin^4(\theta)}{M^2}.
\end{split}
\end{equation}
Substituting Eq.\,\ref{eq:Apendix_S_y0_1} into Eq.\,\ref{eq:Apendix_S_y0}, we obtain
\begin{equation}\label{eq:Apendix_S_y0_2}
\begin{split}
|A^\theta_q(\Omega)-R_q(\Omega)|^2=&|R_q(\Omega)|^2+\frac{\alpha^4\sin^2\theta}{M^2}\\
&+\frac{\alpha^2(\omega_q^2-\Omega^2)\sin(2\theta)}{M}.
\end{split}
\end{equation}
Combining Eq.\,\ref{eq:Apendix_S_y0_2} and Eq.\,\ref{eq:Appendix_S_yq_self}, we obtain:
\begin{equation}
S_{\tilde{y}^q_{\theta}\tilde{y}^q_{\theta}}=\frac{|A^\theta_q(\Omega)-R_q(\Omega)|^2}{|R_q(\Omega)|^2}.
\end{equation}

\section{The effect of classical thermal noise}\label{sec:classical_thermal}
Nonlinear quantum mechanics theory such as Schr{\"o}dinger-Newton theory has a subtle issue when dealing with environmental thermal noise, which was detailed in\,\cite{Helou2017}. In the main text, we have focused on the prescription of thermal noise completely attributed to the quantum origin.  In this section, we follow a different prescription for thermal noise proposed in\,\cite{Helou2017}, where the thermal noise is separated into the classical and the quantum part.
\begin{equation}
\hat{F}_{\rm th}(t)=f_{\rm cl}(t)+\hat{f}_{\rm zp}(t),
\end{equation}
where
\begin{equation}
f_{\rm cl}(t)=\langle\hat{F}_{\rm th}(t)\rangle,\quad \hat{f}_{\rm zp}(t)=\hat{F}_{\rm th}(t)-\langle\hat{F}_{\rm th}(t)\rangle.
\end{equation}
The total thermal noise spectrum is represented by:
\begin{equation}
S_{F_{\rm th}F_{\rm th}}(\Omega)=4\hbar\left[\frac{1}{e^{\frac{\hbar\Omega}{k_BT}}-1}+\frac{1}{2}\right]\frac{{\rm Im}[\chi_m(\Omega)]}{|\chi_m(\Omega)|^2}.
\end{equation}
Using the approach of the paper\,\cite{Helou2017}, the total thermal noise spectrum is separated into the quantum part and classical part as follows,
\begin{equation}
\begin{split}
&S_{f_{\rm cl}f_{\rm cl}}(\Omega)=4\hbar\frac{1}{e^{\frac{\hbar\Omega}{k_BT}}-1}\frac{{\rm Im}[\chi_m(\Omega)]}{|\chi_m(\Omega)|^2}\approx4k_BTM\gamma_m,\\
&S_{f_{\rm zp}f_{\rm zp}}(\Omega)=2\hbar\frac{{\rm Im}[\chi_m(\Omega)]}{|\chi_m(\Omega)|^2}\approx2\hbar\Omega M\gamma_m.
\end{split}
\end{equation}
In the high-temperature limit, we assume the thermal noise is dominantly classical $S_{f_{\rm cl}f_{\rm cl}}(\Omega)\gg S_{f_{\rm zp}f_{\rm zp}}(\Omega)$.

Now, we discuss the effect of classical thermal noise on the self-gravity protocol as an example. In the Heisenberg picture, the classical thermal noise modifies the evolution equation as,
\begin{equation}
\begin{split}
&\dot{\hat{x}}=\frac{\hat{p}}{M},\\
&\dot{\hat{p}}=-M\omega_m^2\hat{x}-M\omega_{SN}^2(\hat{x}-\langle\hat{x}\rangle)+\hbar\alpha\hat{a}_1+f_{\rm cl},\\
&\hat{b}_1=\hat{a}_1,\\
&\hat{b}_\theta=\cos\theta\hat{a}_2+\sin\theta\hat{a}_1+\sin\theta\alpha\hat{x},
\end{split}
\end{equation}
where the outgoing field can be solved as:
\begin{equation}
\begin{split}
&\hat{b}_{\theta}(t)=\cos\theta\hat{a}_1+\sin\theta\hat{a}_2\\
&+\sin\theta\alpha\int_{-\infty}^t\chi_q(t-t')(\hbar\alpha\hat{a}_1(t')+M\omega_{SN}^2\langle\hat{x}(t')\rangle+f_{\rm cl}(t'))dt'.
\end{split}
\end{equation}

Similarly, the thermal noise affect the displacement of mirror $\hat{x}$ in its classical part $x_{\rm cl}$ while preserving its quantum part $\hat x_q$,
\begin{equation}
\begin{split}
\hat{x}_q(t)&=\int_{-\infty}^t\chi_q(t-t')\hbar\alpha\hat{a}_1(t')dt',\\
x_{\rm cl}(t)&=\int_{-\infty}^t\chi_q(t-t')(M\omega_{\rm SN}^2\langle\hat{x}(t')\rangle+f_{\rm cl}(t'))dt'.
\end{split}
\end{equation}
Since the quantum part is unaffected, the Wiener filter that is used to estimate $\langle \hat x_q\rangle$ is unchanged, and thus we have:
\begin{equation}
\langle\hat{x}_q\rangle(t)=\int_{-\infty}^tK_q(t-t')\tilde{y}_\theta^q.
\end{equation}
Transforming into the frequency domain:
\begin{equation}\label{eq:Hei_mater_thermal_1}
\langle\hat{x}(\Omega)\rangle=\frac{\chi_m(\Omega)}{\chi_q(\Omega)}K_q(\Omega)\tilde{y}^q_{\theta}(\Omega)+\chi_m(\Omega)f_{\rm cl}(\Omega).
\end{equation}
Now we can derive the formula for the output data taking into account the classical noise:
\begin{equation}\label{eq:Hei_mater_thermal_2}
\tilde{y}_\theta(\Omega)=\tilde{y}^q_\theta(\Omega)+\chi_q(\Omega)[\sin\theta\alpha M\omega_{SN}^2\langle\hat{x}(\Omega)\rangle+\sin\theta\alpha f_{\rm cl}(\Omega)].
\end{equation}
Using the Wiener filter function,
\begin{equation}
K_q(\Omega)=\frac{A_q^\theta(\Omega)}{\alpha\sin\theta(A_q^\theta(\Omega)-R_q(\Omega))},
\end{equation}
the outgoing result can be derived as:  
\begin{equation}
\begin{split}
\tilde{y}_{\theta}(\Omega)=&\left[\frac{R_q(\Omega}{R_m(\Omega)}\frac{A_q^\theta(\Omega)-R_m(\Omega)}{A_q^\theta(\Omega)-R_q(\Omega)}\right]\tilde{y}_{\theta}^0(\Omega)\\
&+\alpha\sin\theta\chi_m(\Omega)f_{\rm cl}(\Omega),
\end{split}
\end{equation}
therefore, the noise spectrum of the outgoing field becomes:
\begin{equation}
S_{\tilde{y}_\theta\tilde{y}_\theta}=\frac{|A_q^\theta(\Omega)-R_m(\Omega)|^2}{|R_m(\Omega)|^2}+\sin^2\theta\alpha^2|\chi_m(\Omega)|^2S_{f_{\rm cl}f_{\rm cl}}(\Omega).
\end{equation}
Using the expression of $A_q^\theta(\Omega)$ and $R_m(\Omega)$, the noise spectrum can be represented by:
\begin{equation}\label{eq:spectrum_heisenberg_classical_thermal}
\begin{split}
&S_{\tilde{y}_\theta\tilde{y}_\theta}(\Omega)=1+|\chi_m(\Omega)|^2[\alpha^4\sin^4\theta+\alpha^2\sin2\theta(\omega_m^2-\Omega^2)]\\
&+4|\chi_m(\Omega)|^2\sin\theta^2\alpha^2M\gamma_mk_BT+\omega_{\rm SN}^2M^2|\chi_m(\Omega)|^2\times\\
&\left(2\omega_q^2+\Lambda^2\sin2\theta-2\sqrt{(\Lambda^4\sin^2\theta+\omega_q^2+2\Lambda^2\omega_q^2\sin2\theta)}\right).
\end{split}
\end{equation}
The influence of the classical thermal noise on the mutual gravity protocol can be similarly analyzed.

In the main text, Fig.\,\ref{fig:SN_detV} and Fig.\,\ref{fig:entanglement_two_mirrors} depict the effects of classical thermal noise in both SN self-gravity and mutual SN gravity cases. These results indicate that the influence of classical thermal noise closely resembles that observed in quantum gravity scenarios. Specifically, in the self-gravity case, this similarity arises because classical thermal force noise, which is a c-number $f_{\rm cl}$ drives the classical component of $\hat{x}$ in the same way as the quantum thermal noise operator $\hat f_{\rm th}$ drives the quantum displacement operator $\hat x$ in the quantum gravity scenario.
Concretely speaking, we have:
\begin{equation}
\begin{split}
&\hat{x}(\Omega)=\hat{x}_q(\Omega)+x_{\rm cl}(\Omega),\\
&x_{\rm cl}(\Omega)=M\omega_{\rm SN}^2\chi_m(\Omega)K_q(\Omega)\tilde{y}_\theta^q(\Omega)+\chi_m(\Omega)f_{\rm cl}(\Omega),
\end{split}
\end{equation}
where $\hat{x}_q$ is unaffected by classical thermal noise. For comparison, in the case of quantum gravity, the displacement operator of the mirror satisfies:
\begin{equation}
\hat{x}(\Omega)=\chi_m(\Omega)\hbar\alpha\hat{a}_1(\Omega)+\chi_m(\Omega)\hat{f}_{\rm th}(\Omega).
\end{equation}
These two equations demonstrate that the influence of thermal noise manifests in a highly similar manner.  Similarly, in the mutual SN gravity scenario, under the causal conditional framework,  the classical thermal noise \emph{randomly} and \emph{directly} drives the conditional mean of one mirror by influencing the classical component of its $\hat{x}$. The thermal-driven classical stochastic motion of the mirror generates a stochastic gravitational field, which subsequently exerts a random force on the second mirror. Therefore, the optical field reflected from the second mirror motion contains information about the thermal driving of the first mirror. Similar to the self-gravity case, in the mutual SN gravity scenario, the mirror's response to classical thermal noise exhibits a form similar to that observed in the quantum gravity case\,\cite{Miao2020}, except the $\hat f_{\rm th}$ is replaced by $f_{\rm cl}$. However, in the case of a large thermal occupation number,  the $f_{\rm cl}$ and  $\hat{F}_{th}$ follow almost identical statistical distributions. Therefore, if the thermal noise is treated classically, the differences in the correlation spectrum between SN gravity and quantum gravity cases become negligible, as shown in Fig.\,\ref{fig:entanglement_two_mirrors}.

\bibliographystyle{unsrt}
\bibliography{causal-conditional}

\end{document}